# WIGNER FUNCTION PROPERTIES FOR ELECTROMAGNETIC SYSTEMS

**E.E. Perepelkin[a,b,d*], B.I. Sadovnikov[a], N.G. Inozemtseva[b,c], P.V. Afonin[a]**

[a] *Faculty of Physics, Lomonosov Moscow State University, Moscow, 119991 Russia*
[b] *Moscow Technical University of Communications and Informatics, Moscow, 123423 Russia*
[c] *Dubna State University, Moscow region, Dubna,141980 Russia*
[d] *Joint Institute for Nuclear Research, Moscow region, Dubna,141980 Russia*
**Corresponding author: pevgeny@jinr.ru*

## Abstract

Using the Wigner-Vlasov formalism, an exact 3D solution of the Schrödinger equation for a scalar particle in an electromagnetic field is constructed. Electric and magnetic fields are non-uniform. According to the exact expression for the wave function, the search for two types of the Wigner functions is conducted. The first function is the usual Wigner function with a modified momentum. The second Wigner function is constructed on the basis of the Weyl-Stratonovich transform in papers [Phys. Rev. A 35 2791 (1987)] or [Phys. Rev. B 99 014423 (2019)]. It turns out that the second function, unlike the first one, has areas of negative values for wave functions with the Gaussian distribution (Hudson's theorem).

An example of electromagnetic quantum system described by a non-Gaussian wave function has successfully been found. The second Wigner function is positive over the whole phase space for the non-Gaussian wave function. This result is analogous to the Hudson theorem for the gage-invariant Wigner function.

On the one hand, knowing the Wigner functions allows one to find the distribution of the mean momentum vector field and the energy spectrum of the quantum system. On the other hand, within the framework of the Wigner-Vlasov formalism, the mean momentum distribution and the magnitude of the energy are initially known. Consequently, the mean momentum distributions and energy values obtained according to the Wigner functions can be compared with the exact momentum distribution and energy values. This paper presents this comparison and describes the differences.

The Vlasov-Moyal approximation of average acceleration flow has been built in phase space for a quantum system with electromagnetic field. The obtained approximation makes it possible to cut the Vlasov chain off at the second equation and also to analyze the Boltzmann H-function evolution. By averaging of the Vlasov-Moyal approximation over momentum space we can derive an expression for the electromagnetic force of the classical system. This averaging makes the high-order quantum terms disappear from the «motion equation».

## Introduction

The basic concept of the Vlasov theory [1] is presented by distribution function $f_\infty\left(\vec{\xi},t\right)$, defined in infinite-dimensional phase space $\Omega_\infty$ of kinematical quantities of all orders $\vec{\xi}=\left\{\vec{r},\vec{v},\dot{\vec{v}},\ddot{\vec{v}},...\right\}^T\in\Omega_\infty$. Kinematical quantities $\vec{r},\vec{v},\dot{\vec{v}},\ddot{\vec{v}},...$ in a general case are independent variables. Each point $\vec{\xi}_0\in\Omega_\infty$ corresponds to a one-parameter Lie group, which determines the phase trajectory evolution $\vec{\xi}(t)=\exp\left[t\hat{D}\right]\vec{\xi}_0$, where $\hat{D}\vec{\xi}\overset{\text{det}}{=}\vec{u}_\xi\overset{\text{det}}{=}\left\{\vec{v},\dot{\vec{v}},\ddot{\vec{v}},...\right\}^T$ is the tangent velocity vector for trajectory $\vec{\xi}(t)$ in generalized phase space (GPS) $\Omega_\infty$. Trajectory $\vec{\xi}(t)$ actually contains the Taylor series for kinematical quantities $\vec{r},\vec{v},\dot{\vec{v}},\ddot{\vec{v}},...$. In such case kinematical quantities have differential dependencies $\vec{v}=\dot{\vec{r}},\dot{\vec{v}}=\ddot{\vec{r}},...$ and physical system is described by a deterministic motion trajectory $\vec{\eta}(t)$. The distribution function for such a system can be represented in the following form $f_\infty(\vec{\xi},t)=\delta\left(\vec{\xi}-\vec{\eta}(t)\right)$, where $\delta$ is the Dirac delta function. In statistic systems the kinematical quantities $\vec{r},\vec{v},\dot{\vec{v}},\ddot{\vec{v}},...$ are independent, and function $f_\infty$ represents the probability density (normalization condition) or is the distribution function (number of particles). The first Vlasov principle is the probability conservation law in GPS for function $f_\infty$:

$$\frac{\partial}{\partial t}f_\infty+\operatorname{div}_\xi\left(f_\infty\vec{u}_\xi\right)=0, \tag{i.1}$$

where $\operatorname{div}_\xi=\operatorname{div}_r+\operatorname{div}_v+\operatorname{div}_{\dot{v}}+...$. The functional integration of equation (i.1) over phase subspaces $\Omega_n$ GPS $\Omega_\infty=\Omega_1\times\Omega_2\times...$ yields the infinite self-linked Vlasov equation chain:

$$\frac{\partial}{\partial t}f_1+\operatorname{div}_r\left(f_1\langle\vec{v}\rangle\right)=0, \tag{i.2}$$

$$\frac{\partial}{\partial t}f_2+\operatorname{div}_r\left(f_2\vec{v}\right)+\operatorname{div}_v\left(f_2\langle\dot{\vec{v}}\rangle\right)=0, \tag{i.3}$$

$$\frac{\partial}{\partial t}f_3+\operatorname{div}_r\left(f_3\vec{v}\right)+\operatorname{div}_v\left(f_3\dot{\vec{v}}\right)+\operatorname{div}_{\dot{v}}\left(f_3\langle\ddot{\vec{v}}\rangle\right)=0, \tag{i.4}$$

$$....$$

где

$$1=\int_{\mathbb{R}^3}f_1(\vec{r},t)d^3r=\int_{\mathbb{R}^3}\int_{\mathbb{R}^3}f_2(\vec{r},\vec{v},t)d^3rd^3v, \tag{i.5}$$

$$f_1(\vec{r},t)\langle\vec{v}\rangle(\vec{r},t)=\int_{\mathbb{R}^3}f_2(\vec{r},\vec{v},t)\vec{v}d^3v,\quad f_1(\vec{r},t)\langle\langle\dot{\vec{v}}\rangle\rangle(\vec{r},t)=\int_{\mathbb{R}^3}f_2\langle\dot{\vec{v}}\rangle d^3v,$$

$$f_2(\vec{r},\vec{v},t)\langle\dot{\vec{v}}\rangle(\vec{r},\vec{v},t)=\int_{\mathbb{R}^3}f_3(\vec{r},\vec{v},\dot{\vec{v}},t)\dot{\vec{v}}d^3\dot{v}.$$

The physical meaning of equation chain (i.2)-(i.4) becomes transparent when we rewrite it as follows [2]:

$$\hat{\pi}_nS_n=-Q_n,\ n\in\mathbb{N}, \tag{i.6}$$

where

$$\hat{\pi}_n=\frac{\partial}{\partial t}+\vec{v}\nabla_r+\dot{\vec{v}}\nabla_v+...+\left\langle\vec{\xi}^{\,n+1}\right\rangle\nabla_{\xi^n}, \tag{i.7}$$

$$S_n\overset{\text{det}}{=}\operatorname{Ln}f_n,\ Q_n\overset{\text{det}}{=}\operatorname{div}_{\xi^n}\left\langle\vec{\xi}^{\,n+1}\right\rangle. \tag{i.8}$$

Operator (i.7) is the total derivative over time along the phase trajectory in GPS. Quantities $Q_n$ (i.8) define the density of dissipation sources of kinematic values. The initial equation (i.1) does not contain dissipation sources since $Q_\infty = \operatorname{div}_\xi \vec{u}_\xi = 0$. As a result the probability density is constant $f_\infty = const$ along the phase trajectory $\vec{\xi}(t)$ since $\hat{\pi}_\infty f_\infty = \partial_t f_\infty + \vec{u}_\xi \nabla_\xi f_\infty = 0$. The absence of dissipation sources $Q_\infty = 0$ in $\Omega_\infty$ comes from the uniqueness of the Taylor expansions of the trajectory $\vec{\xi}(t) = \exp\left[t\hat{D}\right]\vec{\xi}_0$, that is there is only one generalized phase trajectory matching each point $\vec{\xi}_0 \in \Omega_\infty$. Kinematic information «drops out» with the averaging of (i.5) over phase subspaces $\Omega_n$, resulting in appearance of dissipation sources $Q_n \neq 0$ in equations (i.2)-(i.4) for functions $f_n$ or equations (i.6) for functions $S_n$. Thus, the total derivative over time $\hat{\pi}_n S_n$ along the phase trajectory in phase subspace $\Omega_n$ (i.6) might be not zero. Therefore, a change in probability density $f_n$ along the phase trajectory is due to the presence of dissipation sources $Q_n$. Similar considerations can be applied to the Boltzmann $H_n$ – functions, for which the Vlasov chain yields the following evolutional equations:

$$\hat{\pi}_0\left[f_0 H_n\right] = -f_0\left\langle \ldots \left\langle Q_n \right\rangle \ldots \right\rangle,\ n \in \mathbb{N}, \tag{i.9}$$

$$H_n(t) \overset{\det}{=} \frac{1}{f_0}\int_{\Omega_1}\ldots\int_{\Omega_n} f_n\left(\vec{\xi}^{\,n},t\right) S_n \prod_{s=1}^{n} d^3\xi^s = \left\langle \ldots \left\langle S_n \right\rangle \ldots \right\rangle(t), \tag{i.10}$$

where $f_0$ corresponds to the particle quantity in the system or to the normalization condition for probability density, and $\hat{\pi}_0 = d/dt$. To facilitate understanding for the readers it should be noted that the most known Boltzmann $H-$ function is the $H_2$ function, related to the system entropy as well as the $H-$theorem associated with it. From equations (i.9) it follows that the change in the Boltzmann $H_n$ – function is described by the mean dissipation sources.

Let us consider in details the first two equations from the Vlasov chain (i.2) and (i.3). The mean kinematic quantity $\langle \vec{v} \rangle$ determines the probability flow velocity, and $\langle \dot{\vec{v}} \rangle$ – the probability flow acceleration. The integration over the velocity space of the second Vlasov equation (i.3) results in the first Vlasov equation (i.2). When we multiply the second equation (i.3) by the velocity $v_k$ and integrate it over the velocity space then we get the following motion equation in a hydrodynamic approximation [3]:

$$\hat{\pi}_1\langle v_k \rangle = \left(\frac{\partial}{\partial t} + \langle v_\lambda \rangle \frac{\partial}{\partial x^k}\right)\langle v_k \rangle = -\frac{1}{f_1}\frac{\partial \mathrm{P}_{k\lambda}}{\partial x^k} + \left\langle\left\langle \dot{v}_k \right\rangle\right\rangle, \tag{i.11}$$

$$\mathrm{P}_{k\lambda} = \int_{\Omega_2} f^{1,2}\left(v_k - \langle v_k \rangle\right)\left(v_\lambda - \langle v_\lambda \rangle\right) d^3 v,$$

where $\mathrm{P}_{k\lambda}$ is the pressure tensor. The quantity $m\left\langle\left\langle \dot{v}_k \right\rangle\right\rangle$ stands for the external force, and $-\frac{1}{f_1}\frac{\partial \mathrm{P}_{k\lambda}}{\partial x^k}$ – the pressure force.

Let us keep in mind that the Vlasov chain is self-linked that is to find function $f_1$ we need to know the field $\langle \vec{v} \rangle$, which can be obtained by the function $f_2$ from the second Vlasov equation (i.3) according to (i.5). To solve the Vlasov chain, it is necessary to cut it off at some of its equation. The longer the chain the more kinematic information of the system can be obtained.

To cut the chain off the mean kinematic quantity $\left\langle \vec{\xi}^{\,n+1} \right\rangle$ needs to be approximated. Let us cut the chain off at the first equation. According to the Helmholtz theorem the vector field $\langle \vec{v} \rangle$ allows a decomposition into a potential component $-\alpha \nabla_r \Phi$ and a vortex component $\gamma \vec{A}$

$$\langle \vec{v} \rangle = -\alpha \nabla_r \Phi + \gamma \vec{A}, \tag{i.12}$$

where $\alpha, \gamma$ – are constant values, and $\Phi$, $\vec{A}$ are some functions. Since the probability density $f_1$ is a positive function then $f_1 = |\Psi|^2 \geq 0$, where $\Psi \in \mathbb{C}$. The authors of [4] used decomposition (i.12) from the first Vlasov equation (i.2) and obtained the following equations:

$$\frac{i}{\beta} \frac{\partial \Psi}{\partial t} = -\alpha\beta \left( \hat{\mathrm{p}} - \frac{\gamma}{2\alpha\beta} \vec{A} \right)^2 \Psi + U\Psi, \tag{i.13}$$

$$\Phi(\vec{r},t) \overset{\det}{=} i\,\mathrm{Ln}\left( \frac{\Psi}{\bar{\Psi}} \right) = 2\varphi(\vec{r},t) + 2\pi k,\ k \in \mathbb{Z}, \tag{i.14}$$

$$-\frac{1}{\beta} \frac{\partial \varphi}{\partial t} = -\frac{1}{4\alpha\beta} \left| \langle \vec{v} \rangle \right|^2 + \mathrm{V} \overset{\det}{=} \mathrm{H}, \qquad \mathrm{V} = U + \mathrm{Q}, \quad \mathrm{Q} = \frac{\alpha}{\beta} \frac{\Delta_r |\Psi|}{|\Psi|}, \tag{i.15}$$

$$\hat{\pi}_1 \langle \vec{v} \rangle = \frac{d}{dt} \langle \vec{v} \rangle = -\gamma \left( \vec{E} + \langle \vec{v} \rangle \times \vec{B} \right), \tag{i.16}$$

$$\vec{E} \overset{\det}{=} -\frac{\partial \vec{A}}{\partial t} - \frac{2\alpha\beta}{\gamma} \nabla_r \mathrm{V}, \qquad \vec{B} \overset{\det}{=} \mathrm{curl}_r \vec{A}, \tag{i.17}$$

where $\hat{\mathrm{p}} \overset{\det}{=} -\frac{i}{\beta} \nabla_r$ and $\beta \neq 0$, $\beta \in \mathbb{R}$ is a constant, $U(\vec{r},t) \in \mathbb{R}$ is some function. If the constants $\alpha, \beta, \gamma$ are assumed to be

$$\alpha = -\frac{\hbar}{2m}, \quad \beta = \frac{1}{\hbar}, \quad \gamma = -\frac{q}{m}, \tag{i.18}$$

then equation (i.13) can be transformed into the Schrödinger equation for a scalar particle in electromagnetic field, and equation (i.15) takes a form of the Hamilton-Jacobi equation, and equation (i.16) corresponds to a charged particle motion in electromagnetic field (i.17). Potential $\mathrm{Q}$ (i.15) is known as quantum potential in the de Broglie-Bohm theory of «pilot wave» [5-6]. Scalar potential $\Phi$ (i.14) is actually the phase $\varphi$ of the wave function $\Psi = \sqrt{f_1} \exp(i\varphi)$.

Let us note that from the first Vlasov equation (i.2) we can derive an equation of Pauli and Dirac in the Lorentz gauge [7]. When the systems $qf_1 = \varepsilon_0 \,\mathrm{div}_r \vec{E}$ are self-consistent then the Maxwell system can be constructed [4, 7]. All these results are mathematically rigorous and are based on only one principle, namely the probability conservation law (i.1).

Thus, the first Vlasov equation (i.2) can be used for both classical and quantum systems.

Let us consider the second Vlasov equation (i.3) for function $f_2(\vec{r},\vec{v},t)$ in phase space. Equation (i.3) has been widely used in statistical physics, astrophysics, plasma physics and accelerator physics [8-14]. Equation (i.3), as a rule, is usually utilized in the form (i.6) with dissipation sources $Q_2$

$$\frac{\partial}{\partial t} f_2 + \vec{v}\nabla_r f_2 + \langle \dot{\vec{v}} \rangle \nabla_v f_2 = -f_2 \operatorname{div}_v \langle \dot{\vec{v}} \rangle = -f_2 Q_2. \tag{i.19}$$

To cut off the Vlasov chain on the second equation (i.19) it is necessary to introduce dynamic approximation for vector field $\langle \dot{\vec{v}} \rangle$. Vlasov proposed phenomenologically two approximations for statistical and plasma physics:

$$\langle \dot{\vec{v}} \rangle (\vec{r}, \vec{v}, t) = -\frac{1}{m} \nabla_r U(\vec{r}, t), \tag{i.20}$$

$$\langle \dot{\vec{v}} \rangle (\vec{r}, \vec{v}, t) = \frac{q}{m} \left[ \vec{E}_c(\vec{r}, t) + \vec{v} \times \vec{B}(\vec{r}, t) \right], \tag{i.21}$$

where field $\vec{E}_c$ differs from field $\vec{E}$ from (i.17) by the absence of quantum pressure force $-\nabla_r Q$. Field $\langle \dot{\vec{v}} \rangle$ is determined in the phase space and depends upon velocity (momentum). The right part of the approximation (i.20) does not contain any velocity dependency. The Lorentz force only depends upon velocity in the second approximation (i.21). Both approximations (i.20) and (i.21) obviously have no dissipation sources $Q_2 = 0$. It should be noted that dissipation sources $Q_2 \neq 0$ will have to be taken into consideration for relativistic cases in plasma physics (i.21).

When there are no dissipations sources $Q_2 = 0$ then the second Vlasov equation (i.19) transforms into an analogue of the Liouville equation, conserving the probability density along the phase trajectory.

With no magnetic field influence the second approximation (i.21) transforms into the first approximation (i.20). The averaging of approximation (i.21) over velocity space yields the expression for external electromagnetic force (i.16) without quantum pressure.

Presently there are a lot of literature sources in which both directly and indirectly consider approximations $\langle \dot{\vec{v}} \rangle$ with dissipation sources $Q_2 \neq 0$ [15-19].

The validity of phenomenological approximations $\langle \dot{\vec{v}} \rangle$ (i.20)-(i.21) can be confirmed from the quantum mechanics point of view. As the first Vlasov equation is related with quantum mechanics (i.13)-(i.17), it would be logical to consider such relation for the second equation (i.19). Due to the Heisenberg uncertainty principle the consideration of quantum systems in the phase space may seem to be strange. Nevertheless, in 1932 Wigner and Veil [20-21] proposed phenomenologically a function for quasi-probability density (presently known as the Wigner function) for coherent states and its expansion for mixed states:

$$W(\vec{r}, \vec{p}, t) = \frac{1}{(2\pi\hbar)^3} \int_{\mathbb{R}^3} \left\langle \vec{r} + \frac{\vec{s}}{2} \right| \hat{\rho} \left| \vec{r} - \frac{\vec{s}}{2} \right\rangle e^{-\frac{i}{\hbar}\vec{s}\cdot\vec{p}} d^3 s, \tag{i.22}$$

where $\hat{\rho}$ is the density matrix. The function (i.22) distinguishes with its negative values that explains its name «quasi-probability» density. The Hudson theorem [22] for 1D cases and its generalization for 3D cases [23] state that the function (i.22) is only positive for wave functions with the Gauss distribution.

Using the von Neumann equation for the density matrix or the Schrödinger equation for the wave function we can derive the Moyal equation for function (i.22) without magnetic field [24]

$$\frac{\partial W}{\partial t}+\frac{1}{m}\vec{p}\cdot\nabla_r W-\nabla_r U\cdot\nabla_p W=\sum_{l=1}^{+\infty}\frac{(-1)^l(\hbar/2)^{2l}}{(2l+1)!}U\left(\overleftarrow{\nabla}_r\cdot\vec{\nabla}_p\right)^{2l+1}W. \quad \text{(i.23)}$$

The Moyal equation (i.23) makes it possible to derive the second Vlasov equation by introducing the Vlasov-Moyal approximation [25]

$$f_2\langle\dot{v}_k\rangle=\sum_{l=0}^{+\infty}\frac{(-1)^{l+1}(\hbar/2)^{2l}}{m^{2l+1}(2l+1)!}\frac{\partial U}{\partial x^k}\left(\overleftarrow{\nabla}_r\cdot\vec{\nabla}_v\right)^{2l}f_2. \quad \text{(i.24)}$$

Substituting the approximation (i.24) into the second Vlasov (i.19)/(i.3) we obtain the Moyal equation (i.23) for the Wigner function $f_2(\vec{r},\vec{v},t)=m^3W(\vec{r},\vec{p},t)$. It should be noted that no condition was implied for the distribution function to be positive while obtaining the Vlasov chain (i.1). That is why negative values of the Wigner function do not contradict with the Vlasov chain.

The first summand in approximation (i.24) matches the phenomenological Vlasov approximation (i.20). The subsequent summands have coefficients $\hbar^{2l}$ and depend upon velocity. This fact confirms the left part of expression (i.20). The approximation integration (i.24) over velocity (momentum) space leads to the Vlasov approximation of the external force from the motion equation (i.11)

$$\langle\langle\dot{v}_k\rangle\rangle=-\frac{1}{m}\frac{\partial U}{\partial x^k}. \quad \text{(i.25)}$$

The approximation (i.25) in contrast to (i.24) is independent of quantum corrections represented by terms with $\hbar^{2l}$ coefficients. Since both (i.11) and (i.16) motion equations result from the Vlasov chain, they are equivalent. Comparing equations (i.11), (i.16) and approximation (i.25) in case without magnetic field the following expression can be derived

$$\frac{1}{m}\frac{\partial \mathrm{Q}}{\partial x^k}=\frac{1}{f_1}\frac{\partial \mathrm{P}_{k\lambda}}{\partial x^\lambda}, \quad \text{(i.26)}$$

that is the quantum potential give birth to quantum pressure.

Let us note that exact solutions of the first Vlasov equation (i.2) allow us to obtain exact solutions of the Schrödinger equations (i.13), and the Moyal equations (the second Vlasov equations) and to construct the Wigner functions. The reverse way is also possible to construct solutions for the classical systems by the exact solution of the quantum systems [26-28].

To make sure the approximation (i.21) is valid we have to consider a system with electromagnetic interaction in phase space. There are two ways possible. The first one is to construct an evolution equation for the Wigner function (i.22) taking electromagnetic field into consideration. Such a way was earlier investigated in [29-32]. The function (i.22) appeared to not possess a gauge invariance for electromagnetic fields. Thus, the second way seems to be more promising, which use the Weyl-Stratonovich transformation [33-37] to construct a new Wigner function $f_w$:

$$f_w(\vec{r},\vec{P},t)=\frac{1}{(2\pi\hbar)^3}\int_{\mathbb{R}^3}e^{-\frac{i}{\hbar}\vec{s}\cdot\left[\vec{P}+\frac{q}{2}\int_{-1}^{1}\vec{A}\left(\vec{r}+\tau\frac{\vec{s}}{2},t\right)d\tau\right]}\rho\left(\vec{r}+\frac{\vec{s}}{2},\vec{r}-\frac{\vec{s}}{2},t\right)d^3s, \quad \text{(i.27)}$$

where $\vec{P} = \vec{p} - q\vec{A}$ and $\rho(\vec{r}_+, \vec{r}_-, t) = \bar{\Psi}(\vec{r}_-, t)\Psi(\vec{r}_+, t)$.

At this point several ways come up to proceed with the investigation.

First of all, a comparison seems to be interesting of the Wigner functions (i.22) and (i.27) on exact model solutions for quantum systems with electromagnetic interaction and preferably with non-uniform fields. Such comparison can reveal peculiarities of each function. For instance, the function (i.22) is simpler than the function (i.27). On the other hand, the function (i.27) is gauge invariant, but how crucial this can be for calculating average kinematical quantities is an open question. There is the Hudson theorem that might be useful for the function (i.22) but it is not clear if this theorem is also valid for (i.27)?

Another interesting direction for investigation is to obtain an extended Vlasov-Moyal approximation (i.24) for electromagnetic systems. This type of approximation can help to validate expression (i.21) and its possible quantum corrections. In our paper we are going to clarify the following questions. What influence of the magnetic field is on the quantum pressure? Will the quantum corrections be kept after the averaging of the new approximation over the velocity space? The Vlasov-Moyal approximation for electromagnetic systems will allow average dissipation sources $Q_2$ to be found and the Boltzmann $H_2$– function (i.9) evolution to be estimated. To construct the Vlasov-Moyal approximation taking electromagnetic field into account we need to obtain evolution equations for functions (i.22) and (i.27). Evolution equations known from the literature are quite sophisticated and different from the second Vlasov equation. Therefore, we have to derive another formulation similar to the second Vlasov equation (i.3).

The paper has the following structure. In §1, based on the Wigner-Vlasov formalism, a 3D solution $\Psi$ of the Schrödinger equation (i.13)-(i.17) is constructed from the known distribution density $f_1$ and the vector field of probability flow $\langle \vec{v} \rangle$. Functions $f_1$ and $\langle \vec{v} \rangle$ satisfy the first Vlasov equation (i.2). Using representations (i.22) and (i.27), the distribution functions $W$ and $f_w$ are found. It turns out that in contrast to function $W$, function $f_w$ has regions of negative values for the wave function with the Gaussian distribution (Hudson's theorem).

Moreover, an example has succeeded to be found of the electromagnetic quantum system described by a non-Gauss wave function, which function $f_w$ is positive over the whole phase space. This result resembles the Hudson theorem for the function $f_w$.

Knowing functions $W$ and $f_w$, one can calculate momentum fields $\langle \vec{p} \rangle$, $\langle \vec{P} \rangle$ and the mean value of the energy, compare them with exact distribution and exact value of the energy.

The functions $W$ and $f_w$ are shown to produce different densities of momentum distribution.

In §2, an evolution equation is constructed for the Wigner function $W$, taking into account the electromagnetic interaction. In contrast to the equation for function $f_w$, the resulting equation for function $W$ has a compact notation similar to equation (i.3).

The main difference of the equation constructed for function $W$ from the previous known equations is that it resembles the second Vlasov equation (i.3). This resemblance has allowed the Vlasov-Moyal approximation (i.24) to be extended on the quantum systems with electromagnetic field. Some limit cases of the Vlasov-Moyal approximations have been considered both for classical and quantum systems in details. Quantum corrections in the external forces have been shown to disappear when averaging the Vlasov-Moyal approximation over momentum space. And quantum potential Q in motion equations (i.11), (i.16) has proved to contain information about the quantum nature of the system. If the Vlasov-Moyal approximation is used then the quantum potential Q itself does not explicitly depend upon the magnetic field.

The Appendix section contains proofs of theorems and intermediate mathematical transforms.

## §1 Exact solution of the Schrödinger equation

An analysis of properties of the Wigner functions (i.22) and (i.27) is done using exact model solutions of the Schrödinger equation for quantum systems with electromagnetic field. Previously, as shown in [26], a class of exact solutions ($\Psi$ - model) of the Schrödinger equation (i.13) was constructed, which scalar potential (i.14) had the following form

$$\varphi = k\phi + n\theta + c_0 t,\ \ n,k \in \mathbb{Z} \ \ (or\ n,k \in \mathbb{R}), \tag{1.1}$$

where $\phi,\theta$ is the azimuth and polar angle respectively in the spheric coordinate system, and $c_0 = -\mathrm{E}/\hbar$. Since phase (1.1) is not a continuum function, the potential fraction $\nabla_r \Phi$ in the expansion (i.12) contains a vortex component $\frac{k}{r\sin\theta}\vec{e}_\phi$:

$$\langle \vec{v} \rangle = -2\alpha \nabla_r \varphi = -2\alpha \left( \frac{k}{r\sin\theta}\vec{e}_\phi + \frac{n}{r}\vec{e}_\theta \right). \tag{1.2}$$

Substituting (1.2) in expression (i.17), we can get the following expression

$$\vec{B} = -\frac{q_m^{(Wb)}\delta(\rho)}{2\pi\rho}\vec{e}_z, \quad q_m^{(Wb)} q_e = 2\pi\hbar k,\ \ \rho = r\sin\theta, \tag{1.3}$$

where $q_m^{(Wb)}$ corresponds the magnetic charge and meets the quantization rule [38]. Let us note that magnetic field (1.3) meets the Maxwell classical equation $\operatorname{div}_r \vec{B} = 0$. Intrinsic magnetic moment of a quantum system has the following form $\vec{\mu}_s = \mu_B k \vec{e}_z$, which coincides with the intrinsic magnetic moment of the electron with spin $s = \pm\frac{1}{2}$ when $k = \pm 1$. The solution of equation (i.13) can be represented as

$$\Psi(\vec{r},t) = \frac{1}{\sqrt{\sin\theta}} F_0\left( r, \phi + \frac{k}{n}\left( \operatorname{ctg}\theta - \operatorname{ctg}\left( \theta + \frac{2\alpha n}{r^2}t \right) \right), \theta + \frac{2\alpha n}{r^2}t \right) e^{i\left(n\theta - \frac{\mathrm{E}}{\hbar}t\right)}, \tag{1.4}$$

$$f(r,\phi,\theta,t)\Big|_{t=0} = \frac{F_0(r,\phi,\theta)}{\sin\theta} = f_0(r,\phi,\theta), \tag{1.5}$$

where $F_0$ is some function, defining the class of solutions of the $\Psi$ - model. The function $F_0$ (1.4) is constant along the characteristic curves illustrated in Fig. 1. These characteristic curves in Fig. 1 are given for three different radius $r$ values at $k = 1,\ n = 4$.

Let us consider a private case of decomposition (1.1) with $n = 0,\ \theta = \pi/2$ for a time-independent function $F_0$:

$$f(\vec{r}) = \frac{1}{(2\pi)^{3/2}\sigma_r^3} \exp\left[ -\frac{r^2\sin^2\theta + z^2}{2\sigma_r^2} \right], \tag{1.6}$$

$$\langle \vec{p}_c \rangle = m\langle \vec{v} \rangle = \frac{\hbar}{2}\nabla_r \Phi = \hbar\nabla_r \varphi = \hbar\nabla_r \phi = -q\vec{A} = \frac{\hbar}{\rho}\vec{e}_\phi. \tag{1.7}$$

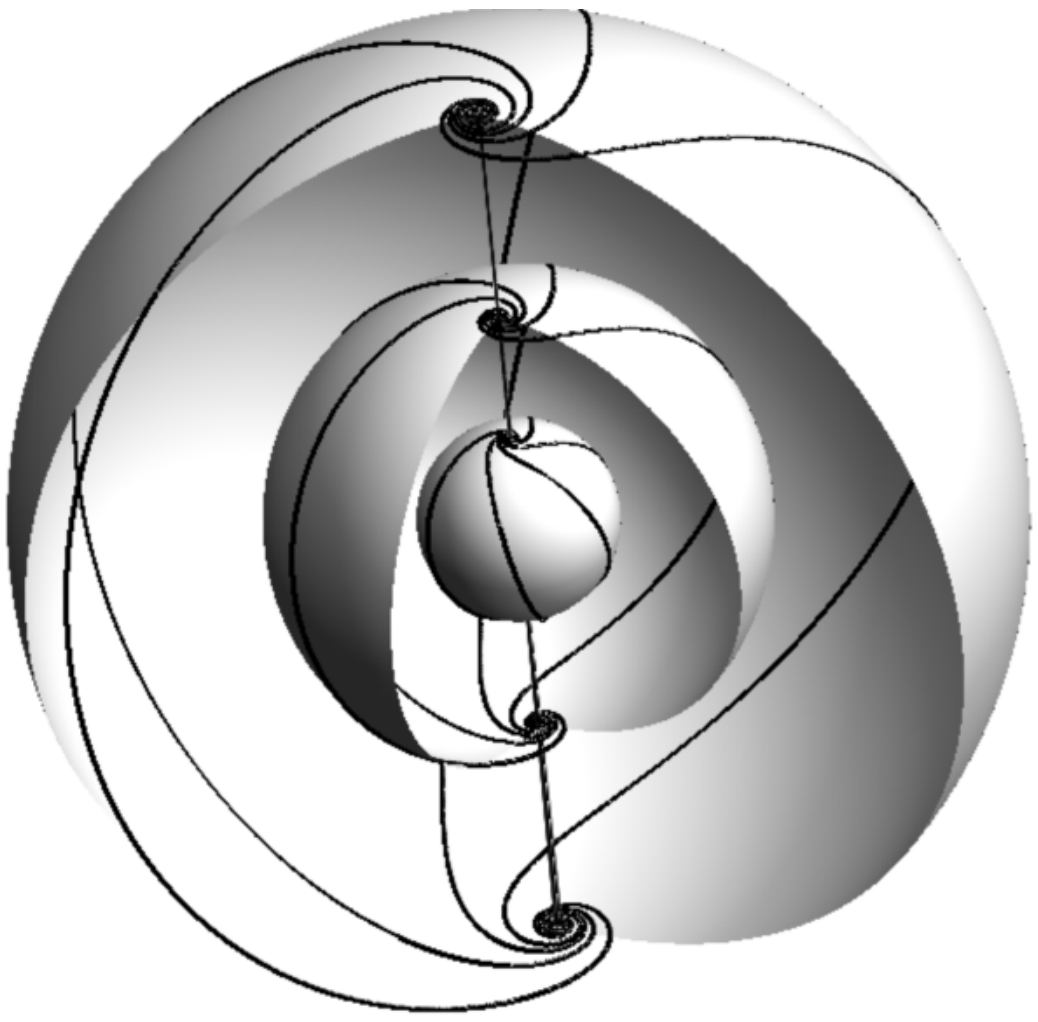

Fig. 1 Characteristic curves

Constant value $\sigma_r$ is a free parameter. Function (1.6) satisfies the normalization condition $\int f\, d^3 r = 1$ and the first Vlasov equation. Thus, solutions (1.6)-(1.7) of the first Vlasov equation (i.2) correspond to two quantum systems. The first system has only electrical interaction $\langle \vec{p}_c \rangle = \hbar \nabla_r \varphi$, and the second – only electromagnetic one $\langle \vec{p}_c \rangle = -q\vec{A}$.

As it will be shown below the both systems are gauge invariant and allow the both Wigner functions (i.22) and (i.27) to be compared.

The following theorem is true.

**Theorem 1** *Wave functions*

$$\Psi^{(E)}(\rho,\phi,z,t) = \frac{1}{(2\pi)^{3/4}\sigma_r^{3/2}} \exp\left(-\frac{\rho^2+z^2}{4\sigma_r^2} + i\phi - i\frac{E}{\hbar}t\right), \tag{1.8}$$

$$\Psi^{(EM)}(\vec{r},t) = \frac{1}{(2\pi)^{3/4}\sigma_r^{3/2}} \exp\left(-\frac{r^2}{4\sigma_r^2} - i\frac{E}{\hbar}t\right), \tag{1.9}$$

*are the exact solutions of the Schrödinger equations:*

$$i\hbar\frac{\partial \Psi^{(E)}}{\partial t} = \frac{\hat{p}^2}{2m}\Psi^{(E)} + U_1\Psi^{(E)}, \tag{1.10}$$

$$i\hbar\frac{\partial \Psi^{(EM)}}{\partial t} = \frac{1}{2m}\left(\hat{p} - q\vec{A}_1\right)^2\Psi^{(EM)} + U_1\Psi^{(EM)}, \tag{1.11}$$

*where*

$$U_1(\vec{r}) = \frac{\hbar^2}{8m\sigma_r^4}\left(r^2 - \frac{4\sigma_r^4}{r^2\sin^2\theta}\right), \qquad q\vec{A}_1(\vec{r}) = -\frac{\hbar}{r\sin\theta}\vec{e}_\phi, \qquad E = \frac{3\hbar^2}{4m\sigma_r^2}, \tag{1.12}$$

*with that, for both systems (1.8) and (1.9) the Hamilton-Jacobi equation (i.15) is satisfied:*

$$-\hbar\frac{\partial\varphi}{\partial t} = \frac{1}{2m}\left|\langle \vec{p}_c \rangle\right|^2 + V, \quad V = U + Q, \tag{1.13}$$

*where the quantum potential (i.15) has the form:*

$$Q(\vec{r}) = \frac{\hbar^2}{4m\sigma_r^2}\left(3 - \frac{r^2}{2\sigma_r^2}\right). \tag{1.14}$$

The proof of Theorem 1 is given in Appendix A.

Fig. 2 shows the graph of potential $U_1$ (1.12) at $z=0$. Potential $U_1$ has the shape of a «quadratic funnel». In the vicinity of axis $OZ$ there is second-order pole $U_1 \sim 1/\rho^2$, and at infinity there is a quadratic dependence $U_1 \sim r^2$. In the absence of a magnetic field, solution $\Psi^{(\mathrm{EM})}$ corresponds to the wave function of the ground state of a harmonic oscillator with quadratic potential $U_1 \sim r^2$.

Let us construct a solution of the Schrödinger equation (1.11) for a modified field (1.7)

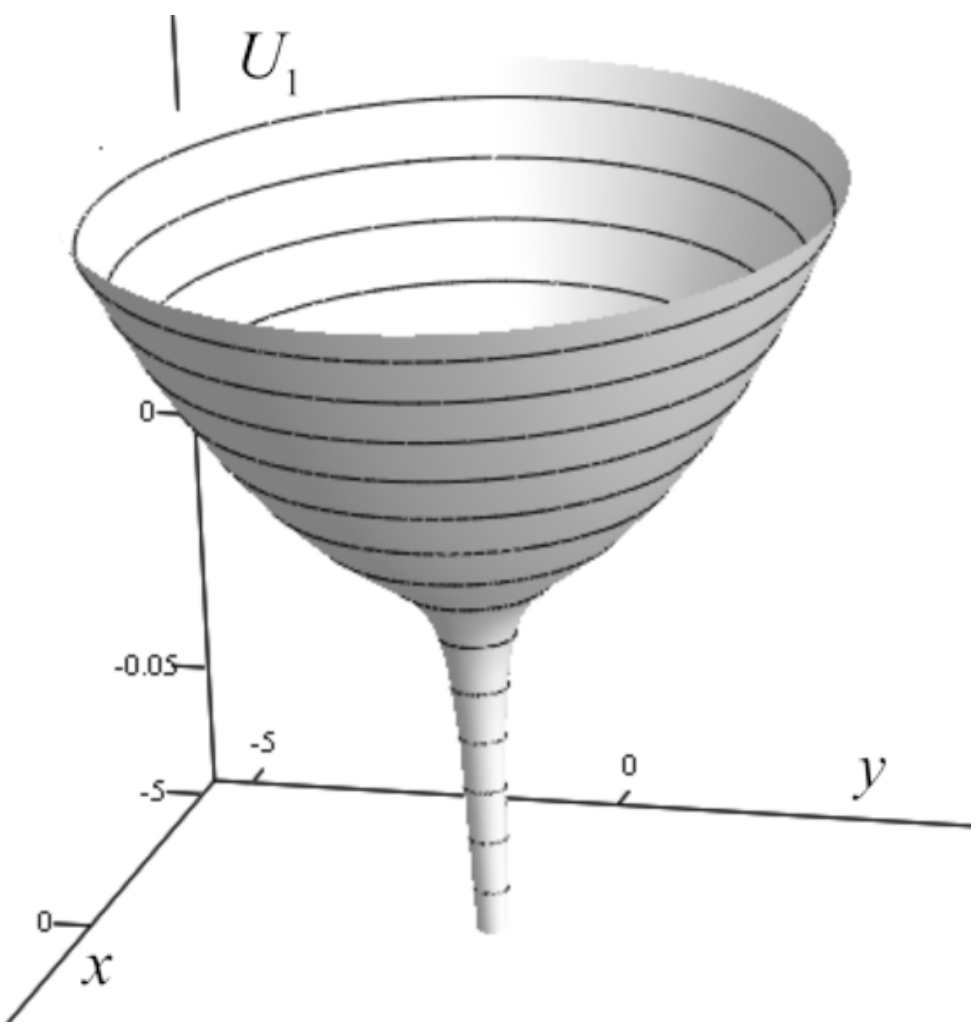

Fig. 2 «Quadratic funnel» potential

$$\langle \vec{p}_c \rangle = m\langle \vec{v} \rangle = -q\vec{A}_2 = \frac{\hbar\eta}{2\sigma_r^2}\rho\vec{e}_\phi, \tag{1.15}$$

where $\eta \in \mathbb{R}$ is some coefficient. The potential (1.15) determines a homogeneous magnetic field $\vec{B}$, which takes the following form at $\eta = k \in \mathbb{Z}$ (see in comparison with (1.3)):

$$\vec{B}_2 = -\frac{q_m^{(Wb)}\eta}{2\pi\sigma_r^2}\vec{e}_z. \tag{1.16}$$

We can propose the following theorem.

**Theorem 2** *Wave function (1.9) is an exact solution of the Schrödinger equation (1.11) with vector potential (1.15), quantum potential (1.14) and potential energy*

$$U_2(\vec{r}) = \frac{\hbar^2}{8m\sigma_r^4}\left[\left(1-\eta^2\right)\rho^2 + z^2\right]. \tag{1.17}$$

The proof of Theorem 2 is given in Appendix A.

In contrast to potential energy $U_1$ (1.12) the expression for $U_2$ (1.17) does not have a pole on the $OZ$ axis. The expression (1.17) becomes independent from polar radius $\rho$ at $\eta = 1$. The quantum system (1.9) is «confined» along the radial direction $\vec{e}_\rho$ at the expense of magnetic field (1.16). The coefficient $1-\eta^2$ is negative in the expression (1.17) at $\eta > 1$. The expansion of quantum system (1.9) with potential $U_2$ in radial direction $\vec{e}_\rho$ is compensated by reinforcement of the external magnetic field (1.16). A similar situation occurs at $0 < \eta < 1$, when the redundant confinement with potential (1.17) is compensated by weakening the external magnetic field (1.16). Thus, any change of coefficient $\eta$ does not affect the balance between the potential energy (1.17) and the vector potential (1.16), which keep the quantum system balanced and the wave function (1.9) unchanged. The quantum pressure (1.14) stays unchanged at any $\eta$ value.

Let us prove the following theorem about the Wigner function (i.22), which will be used to analyze the properties of the Wigner functions (i.22) and (i.27) on solutions (1.8), (1.9) with potentials (1.12) and (1.17).

**Theorem 3** *If the vector field of probability flow* $\langle \vec{v} \rangle$ *is potential and allow the Helmholtz decomposition (i.12) to be made*

$$\langle \vec{v} \rangle = -\alpha \nabla_r \Phi, \tag{1.18}$$

*then function* $f_2(\vec{r}, \vec{p}, t)$ *can be represented as*

$$f_2(\vec{r}, \vec{p}, t) = \frac{1}{(2\pi\hbar)^3} \int_{\mathbb{R}^3} \Psi(\vec{r}_+, t) \bar{\Psi}(\vec{r}_-, t) e^{-i\frac{\vec{p}\cdot\vec{s}}{\hbar}} d^3 s, \tag{1.19}$$

*where* $\vec{r}_\pm = \vec{r} \pm \vec{s}/2$.

The proof of Theorem 3 is given in Appendix A.

The form of function (1.19) totally coincides with the known Wigner function (i.22) at the coherent state. It should be noted that the Wigner function was initially obtained by a phenomenological method. And now we have managed to derive it as a result of the Helmholtz decomposition (1.18) from Theorem 3.

Knowing the expressions for wave functions (1.8)-(1.9), allows us to find the Wigner functions. The classical Wigner function (i.22), corresponds to wave function $\Psi^{(\mathrm{E})}$. This classical Wigner function was obtained in [25] in the form:

$$W^{(\mathrm{E})}(\rho, z, \vec{p}) = \frac{\rho^2}{2\pi^4 \hbar^3 \sigma_r^2} e^{-\frac{\rho^2 + z^2}{2\sigma_r^2} - \frac{2\sigma_r^2}{\hbar^2} p_z^2} \int_0^{2\pi} d\phi_s' \int_0^{+\infty} e^{-\rho^2 \frac{\rho_s'^2}{2\sigma_r^2} + i\vartheta(\rho, p_\rho, p_\phi, \rho_s', \phi_s')} \rho_s' d\rho_s', \tag{1.20}$$

where

$$\vartheta(\rho, p_\rho, p_\phi, \rho_s', \phi_s') = \mathrm{arctg}(k_1 \sin\phi_s') - 2\frac{\rho\rho_s'}{\hbar}(p_\rho \cos\phi_s' + p_\phi \sin\phi_s'),\ k_1(\rho_s') = \frac{2\rho_s'}{1-\rho_s'^2}.$$

The integral in expression (1.20) cannot be taken explicitly, since it reduces to an elliptic integral. Note that function (1.20) satisfies the Moyal equation (i.23).

For wave function $\Psi^{(\mathrm{EM})}$, two Wigner functions can be found: classical function $W^{(\mathrm{EM})}$ (i.22) and function $f_w$ with gauge invariance (i.27). Function $W^{(\mathrm{EM})}$ can be found explicitly (see Appendix C):

$$W^{(\mathrm{EM})}(\vec{r}, \vec{p}) = \frac{1}{(\pi\hbar)^3} \exp\left(-\frac{r^2}{2\sigma_r^2} - \frac{2\sigma_r^2}{\hbar^2} p^2\right). \tag{1.21}$$

Note that function $W^{(\mathrm{EM})}$ does not satisfy the Moyal equation (i.23), since it was obtained for the Schrödinger equation (1.11). In §2, we construct an analogue of the Moyal equation (i.23) for systems with electromagnetic interaction.

To find the function $f_w$, we need the following lemma.

**Lemma 1** *Let vector potential* $q\vec{A}_1(\vec{r}) = -\frac{\hbar}{r\sin\theta}\vec{e}_\phi$ *(1.12), then the expression for the Wigner function* $f_w$ *(i.27) of the system (1.9) can be represented as*:

$$f_w(\vec{r},\vec{P}) = \frac{1}{(2\pi\hbar)^3}\frac{e^{-\frac{r^2}{2\sigma_r^2}}}{(2\pi)^{3/2}\sigma_r^3}\int_{\mathbb{R}^3} e^{-\frac{s^2}{8\sigma_r^2}-\frac{i}{\hbar}\vec{s}\cdot\vec{P}} e^{i(\phi_+-\phi_-)} d^3 s, \tag{1.22}$$

*where* $\phi_\pm = \operatorname{arctg}\frac{y \pm y_s/2}{x \pm x_s/2}$, $\vec{s} = \{x_s, y_s, z_s\}$.

The proof of the lemma 1 is given in Appendix A.

Using the assertion of the lemma, we can prove the following theorem.

**Theorem 4** *Wigner function* $f_w$ *for the quantum system (1.11)-(1.12) with electromagnetic interaction coincides with the Wigner function* $W^{(\mathrm{E})}$ *(1.20) for the quantum system (1.10), (1.12) only with the electric interaction, i.e.*

$$f_w(\vec{r},\vec{P}) = W^{(\mathrm{E})}(\rho, z, \vec{P}). \tag{1.23}$$

The proof of Theorem 4 is given in Appendix A.

**Remark**

From a physical point of view, statement (1.23) of Theorem 4 shows the «equivalence» of two quantum systems (1.8) and (1.9) having the same probability quasi-density function (1.20). This result is expected, since both systems are obtained from the same initial distributions (1.6) and (1.7) in the framework of the Wigner-Vlasov formalism. The difference between distributions $f_w$ and $W^{(\mathrm{E})}$ is the replacement of «electrical» momentum $\vec{p}$ with electromagnetic momentum $\vec{P}$.

From the mathematical point of view the wave functions (1.8), (1.9) and the associated Schrödinger equations (1.10), (1.11) are interrelated with the gauge invariance. The following is actually true

$$\Psi^{(\mathrm{EM})}(\vec{r},t) = \Psi^{(\mathrm{E})}(\vec{r},t)\exp(-i\phi), \tag{1.24}$$

where

$$-\nabla_r \phi = -\frac{1}{r\sin\theta}\vec{e}_\phi = \frac{q}{\hbar}\vec{A}_1(\vec{r}), \qquad \frac{\partial\phi}{\partial t} = 0. \tag{1.25}$$

Therefore, an electromagnetic quantum system with potentials $U_1/q, \vec{A}_1(\vec{r})$ is only invariant with respect to a system with electric field $U_1 - \frac{\partial\phi}{\partial t} = U_1$. Since function $f_w$ possesses the gauge invariance it has the same form (1.22) and (1.23) for both representations (1.8) and (1.9).

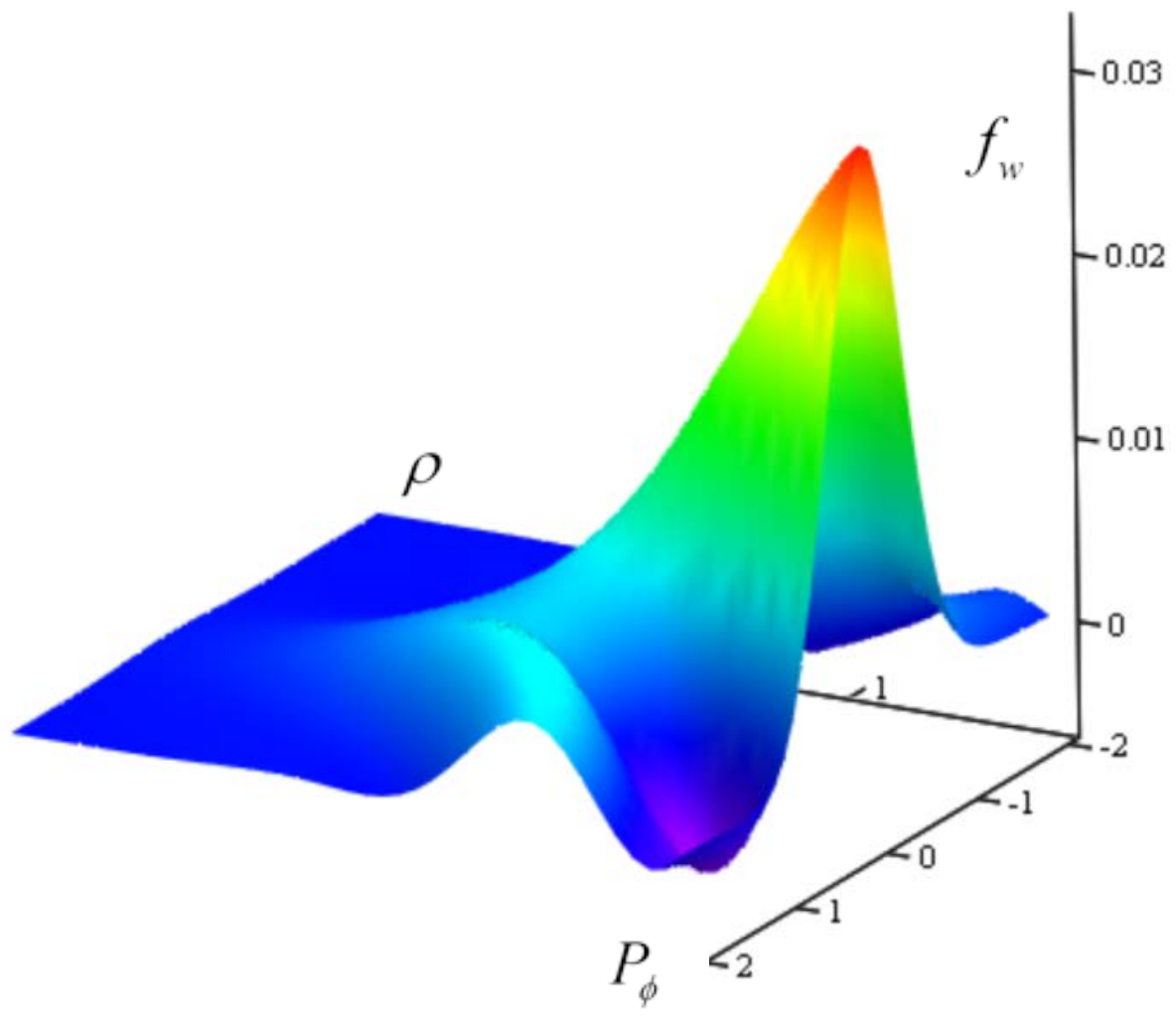

Fig. 3 Regions of negative values of function $f_w$

**Corollary 1** *Wigner function for a quantum system with electromagnetic interaction $f_w$ (i.27) has regions of negative values for the wave function with the Gaussian distribution (1.9).*

As is known, the Wigner function (i.22) is a quasi-density of probabilities of a quantum system, since it has regions of negative values [20, 21]. According to the Hudson's theorem [22] and its generalization for 3D systems [23], the Wigner function (i.22) is positive only for a wave function with the Gaussian distribution. Expression (1.20) for function $f_w(\vec{r},\vec{P}) = W^{(E)}(\rho, z, \vec{P})$ was obtained for the wave function (1.9) in the form of the Gaussian distribution, but function $f_w$ has regions of negative values (see Fig. 3). The regions of negative values of function $W^{(E)}$ are described in [25]. Fig. 3 shows the distribution of function $f_w$ in plane $(\rho, P_\phi)$ at $z=0$, $P_\rho = P_z = 0$.

**Corollary 2** *Wave function (1.8) is a solution for the Schrödinger electromagnetic equation*

$$i\hbar\frac{\partial\Psi}{\partial t} = \frac{1}{2m}\left(\hat{\mathrm{p}} - q\vec{A}_3\right)^2\Psi + U_1\Psi, \qquad q\vec{A}_3(\vec{r}) = \frac{\hbar}{r\sin\theta}\vec{e}_\phi, \tag{1.26}$$

*and the Wigner function $f_w$, associated with it, is positive within the whole phase space and have the following form:*

$$f_w(r,P) = \frac{1}{(\pi\hbar)^3}\exp\left(-\frac{r^2}{2\sigma_r^2} - \frac{2\sigma_r^2}{\hbar^2}P^2\right). \tag{1.27}$$

The proof of Corollary 2 is given in Appendix A.

Let us note that the magnetic field of quantum system (1.26) differs from that of system (1.12) only by its sign.

Corollary 2 is analogous to the Hudson theorem for the Wigner function $f_w$. Unlike the usual Wigner function $W$ (i.22), which is only positive for the Gauss distribution, the function $f_w$ (i.27) is positive for the electromagnetic system (1.26), described by non-Gaussian wave function (1.8). According to Corollary 1 the function $f_w$ has negative regions for the Gaussian wave function.

Let us note that the positiveness of function $f_w$ reduces to the positiveness of the Wigner function W with the following expression under the integral sign:

$$\mathrm{W}(\vec{r},\vec{P},t) = \frac{1}{(2\pi\hbar)^3}\int_{\mathbb{R}^3} e^{-\frac{i}{\hbar}\vec{s}\cdot\vec{P}}\rho_{new}(\vec{r}_+,\vec{r}_-,t)\,d^3s = f_w(\vec{r},\vec{P},t), \tag{1.28}$$

$$\rho_{new}\left(\vec{r}_+,\vec{r}_-,t\right)=\rho\left(\vec{r}_+,\vec{r}_-,t\right)\exp\left[-i\frac{q}{\hbar}\int_{\vec{r}_-}^{\vec{r}_+}\vec{A}\left(\vec{r}',t\right)d\vec{r}'\right], \tag{1.29}$$

where the integral term in (1.29) is taken along the straight line connecting the points $\vec{r}_\pm=\vec{r}\pm\vec{s}/2$. According to the Hudson theorem and its generalization for the 3D case, the quantity (1.29) must conform to the Gauss distribution. Since the positive Wigner function (i.22) has the same form as (1.27), the expression $\rho_{new}\left(\vec{r}_+,\vec{r}_-,t\right)$ must also conform to the Gauss distribution (the Fourier transform of the Gauss function is the Gauss function itself). Expressions (1.8) and (1.27) reduce $\rho_{new}\left(\vec{r}_+,\vec{r}_-,t\right)$ to the Gauss distribution (Corollary 2). We should keep in mind that quantum system (1.26) is a private case of the $\Psi$ - model.

Let us calculate the mean values of momenta $\left\langle\vec{p}\right\rangle$ and $\left\langle\vec{P}\right\rangle$ from the Wigner functions $f_w$ and $W^{(\mathrm{EM})}$. In [25], $\left\langle\vec{p}\right\rangle$ was found from function $W^{(\mathrm{E})}$:

$$\left\langle\vec{p}\right\rangle\left(\vec{r}\right)=\frac{1}{\left|\Psi^{(\mathrm{E})}\right|^2}\int_{\mathbb{R}^3}\vec{p}W^{(\mathrm{E})}\left(\vec{r},\vec{p}\right)d^3p=-q\vec{A}\left(\vec{r}\right)=\left\langle\vec{p}_c\right\rangle\left(\vec{r}\right). \tag{1.30}$$

It follows from result (1.30) and Theorem 4 (1.23) that:

$$\left\langle\vec{P}\right\rangle\left(\vec{r}\right)=\frac{1}{\left|\Psi^{(\mathrm{EM})}\right|^2}\int_{\mathbb{R}^3}\vec{P}f_w\left(\vec{r},\vec{P}\right)d^3P=\frac{1}{\left|\Psi^{(\mathrm{E})}\right|^2}\int_{\mathbb{R}^3}\vec{P}W^{(\mathrm{E})}\left(\vec{r},\vec{P}\right)d^3P=\left\langle\vec{p}_c\right\rangle\left(\vec{r}\right). \tag{1.31}$$

In [36, 37], electromagnetic momentum $\vec{P}$ is represented as $\vec{P}=\vec{p}-q\vec{A}$, therefore, $\left\langle\vec{P}\right\rangle=\left\langle\vec{p}\right\rangle-q\vec{A}=-q\vec{A}$, hence $\left\langle\vec{p}\right\rangle=0$. A similar result $\left\langle\vec{p}\right\rangle=0$ is obtained when averaging over function $W^{(\mathrm{EM})}$:

$$\left\langle\vec{p}\right\rangle\left(\vec{r}\right)=\frac{1}{\left|\Psi^{(\mathrm{EM})}\right|^2}\int_{\mathbb{R}^3}\vec{p}W^{(\mathrm{EM})}\left(\vec{r},\vec{p}\right)d^3p=0. \tag{1.32}$$

Therefore, for a system with electromagnetic interaction, the mean momentum is determined by $\left\langle\vec{P}\right\rangle$, and for a system with only electrical interaction, it is $\left\langle\vec{p}\right\rangle$, and both of these values coincide with the exact value of $\left\langle\vec{p}_c\right\rangle$ (1.7). For the electromagnetic system (1.9), the vortex field (1.7) is caused by magnetic field $\left\langle\vec{P}\right\rangle=-q\vec{A}$, and the potential part is vortex-free ( $\left\langle\vec{p}\right\rangle=0$). The system (1.8) has no magnetic field $\vec{A}=\vec{\theta}$, so the vortex field (1.7) is determined by the potential part $\nabla_r\varphi$ (the phase of the wave function), i.e. $\left\langle\vec{p}\right\rangle=\frac{\hbar}{\rho}\vec{e}_\phi$.

Note that averaging $\vec{P}=\vec{p}-q\vec{A}$ over function $W^{(\mathrm{EM})}$ gives the correct result:

$$\left\langle\vec{P}\right\rangle\left(\vec{r}\right)=\frac{1}{\left|\Psi^{(\mathrm{EM})}\right|^2}\int_{\mathbb{R}^3}\vec{P}W^{(\mathrm{EM})}\left(\vec{r},\vec{p}\right)d^3p=\frac{1}{\left|\Psi^{(\mathrm{EM})}\right|^2}\int_{\mathbb{R}^3}\vec{p}W^{(\mathrm{EM})}\left(\vec{r},\vec{p}\right)d^3p-q\vec{A}=-q\vec{A}. \tag{1.33}$$

Let us calculate energy of state $\mathrm{E}$ (1.12). For both the Wigner functions $f_w$ and $W^{(\mathrm{EM})}$ we are going to average the expression $\mathcal{E}(\vec{r},\vec{P}) = \frac{P^2}{2m} + U_1(\vec{r})$. Direct calculations give the following result (see Appendix C):

$$\begin{aligned} \langle\langle \mathcal{E} \rangle\rangle &= \int\limits_{\mathbb{R}^3}\int\limits_{\mathbb{R}^3}\left[\frac{1}{2m}\left(\vec{p}-q\vec{A}\right)^2 + U_1\right] W^{(\mathrm{EM})}(\vec{r},\vec{p})\, d^3r d^3p = \\ &= \int\limits_{\mathbb{R}^3}\int\limits_{\mathbb{R}^3}\left(\frac{P^2}{2m}+U_1\right) f_w(\vec{r},\vec{P})\, d^3r d^3P = \mathrm{E} = \frac{3\hbar^2}{4m\sigma_r^2}. \end{aligned} \tag{1.34}$$

It follows from expression (1.34) that energy $\mathcal{E}$ averaged $\langle\langle \mathcal{E} \rangle\rangle$ over both the Wigner functions $f_w$ and $W^{(\mathrm{EM})}$ has the same value, which coincides with the energy of quantum system $\mathrm{E}$ (1.12).

It should be noted that the results (1.31)-(1.33) can be expanded to the general case of functions $f_w(\vec{r},\vec{P})$ and $W(\vec{r},\vec{p},t)$. The following lemma can be formulated.

**Lemma 2** *The average momentum field* $\langle \vec{P} \rangle$ *calculated by the Wigner functions* $f_w(\vec{r},\vec{P},t)$ *or* $W(\vec{r},\vec{p},t)$ *can be represented as the Helmholtz decomposition (i.12), (i.14):*

$$\langle \vec{P} \rangle(\vec{r},t) = \frac{1}{|\Psi|^2}\int\limits_{\mathbb{R}^3}\vec{P} f_w(\vec{r},\vec{P},t)\, d^3P = \frac{1}{|\Psi|^2}\int\limits_{\mathbb{R}^3}\vec{P} W(\vec{r},\vec{p},t)\, d^3p = \frac{\hbar}{2}\nabla_r\Phi - q\vec{A}, \tag{1.35}$$

*or*

$$\vec{J}(\vec{r},t) = |\Psi|^2\langle \vec{P} \rangle = -\frac{i\hbar}{2}\left(\bar{\Psi}\nabla_r\Psi - \Psi\nabla_r\bar{\Psi}\right) - q\bar{\Psi}\vec{A}\Psi. \tag{1.36}$$

*where* $\vec{P} = \vec{p} - q\vec{A}$.

The proof of the lemma 2 is given in Appendix A.

Let us compare the momentum representations for quantum systems (1.8), (1.9) and (1.17). The following expressions are known to be valid for the Wigner function (i.22):

$$\int\limits_{\mathbb{R}^3} W(\vec{r},\vec{p},t)\, d^3p = \left|\Psi(\vec{r},t)\right|^2, \quad \int\limits_{\mathbb{R}^3} W(\vec{r},\vec{p},t)\, d^3r = \left|\tilde{\Psi}(\vec{p},t)\right|^2, \tag{1.37}$$

where $\tilde{\Psi}$ is the momentum representation of the wave function ($\tilde{\Psi} = \mathcal{F}[\Psi]$ is the Fourier transform of wave function $\Psi$). Would the following expressions (1.37) be valid for function $f_w(\vec{r},\vec{P})$?

**Corollary 3** *For quantum systems (1.9) and (1.15), (1.17) function* $f_w(\vec{r},\vec{P})$ *has the following properties:*

$$\int_{\mathbb{R}^3} f_w\left(\vec{r},\vec{P},t\right) d^3P = \left|\Psi\left(\vec{r},t\right)\right|^2, \tag{1.38}$$

*for an arbitrary vector potential* $\vec{A}$. *The momentum probability density for* $\vec{A}_1$ *(1.12) and* $\vec{A}_2$ *(1.15) has the following form respectively:*

$$\mathrm{F}_w\left(\vec{P}\right) = \int_{\mathbb{R}^3} f_w\left(\vec{r},\vec{P}\right) d^3r,$$

$$\mathrm{F}_w^1\left(\vec{P}\right) = \left|\tilde{\Psi}^{(\mathrm{E})}\right|^2 = \frac{32\sigma_r^5}{\pi\left(2\pi\right)^{3/2}\hbar^5} P_\rho^2 e^{-2\frac{\sigma_r^2}{\hbar^2}P_z^2} \left(\int_0^{\pi/2} e^{-\frac{\sigma_r^2}{\hbar^2}P_\rho^2\cos^2\phi} \cos^2\phi \, d\phi\right)^2, \tag{1.39}$$

$$\mathrm{F}_w^2\left(\vec{P}\right) = \frac{2\sigma_r^2 e^{-\frac{2\sigma_r^2}{\hbar^2}P^2}}{\pi^2\hbar^3\left(1+\eta^2\right)} \left[\sigma_r\sqrt{2\pi} - \frac{4\sigma_r^2\eta}{\hbar\sqrt{1+\eta^2}} P_\rho \int_0^{\pi/2} e^{\frac{2\sigma_r^2\eta^2P_\rho^2\sin^2\phi}{\hbar^2\left(1+\eta^2\right)}} \operatorname{erf}\left(-\frac{\sigma_r\eta\sqrt{2}}{\hbar\sqrt{1+\eta^2}} P_\rho \sin\phi\right) \sin\phi \, d\phi\right], \tag{1.40}$$

*that is*

$$\mathrm{F}_w\left(\vec{P}\right) \neq \left|\tilde{\Psi}^{(\mathrm{EM})}\left(\vec{P}\right)\right|^2 = \left(\frac{\sigma_r\sqrt{2\pi}}{\pi\hbar}\right)^3 e^{-\frac{2\sigma_r^2}{\hbar^2}P^2} = \mathrm{F}\left(\vec{P}\right). \tag{1.41}$$

The proof of Corollary 3 is given in Appendix A.

Thus, the first relationship in (1.37) is true, and the second one is not true because of (1.41). The wave function $\Psi^{(\mathrm{EM})}$ (1.9) is the same for the both quantum systems with vector potentials $\vec{A}_1$ and $\vec{A}_2$. The momentum representation $\tilde{\Psi}^{(\mathrm{EM})}$ (1.41) of the wave function (1.9) is also the same for the both quantum system. Consequently, the probability density $\left|\tilde{\Psi}^{(\mathrm{EM})}\right|^2$ (1.41) is the same for the both systems. And the same probability density result from the Wigner $W$ (i.22) function according to (1.37).

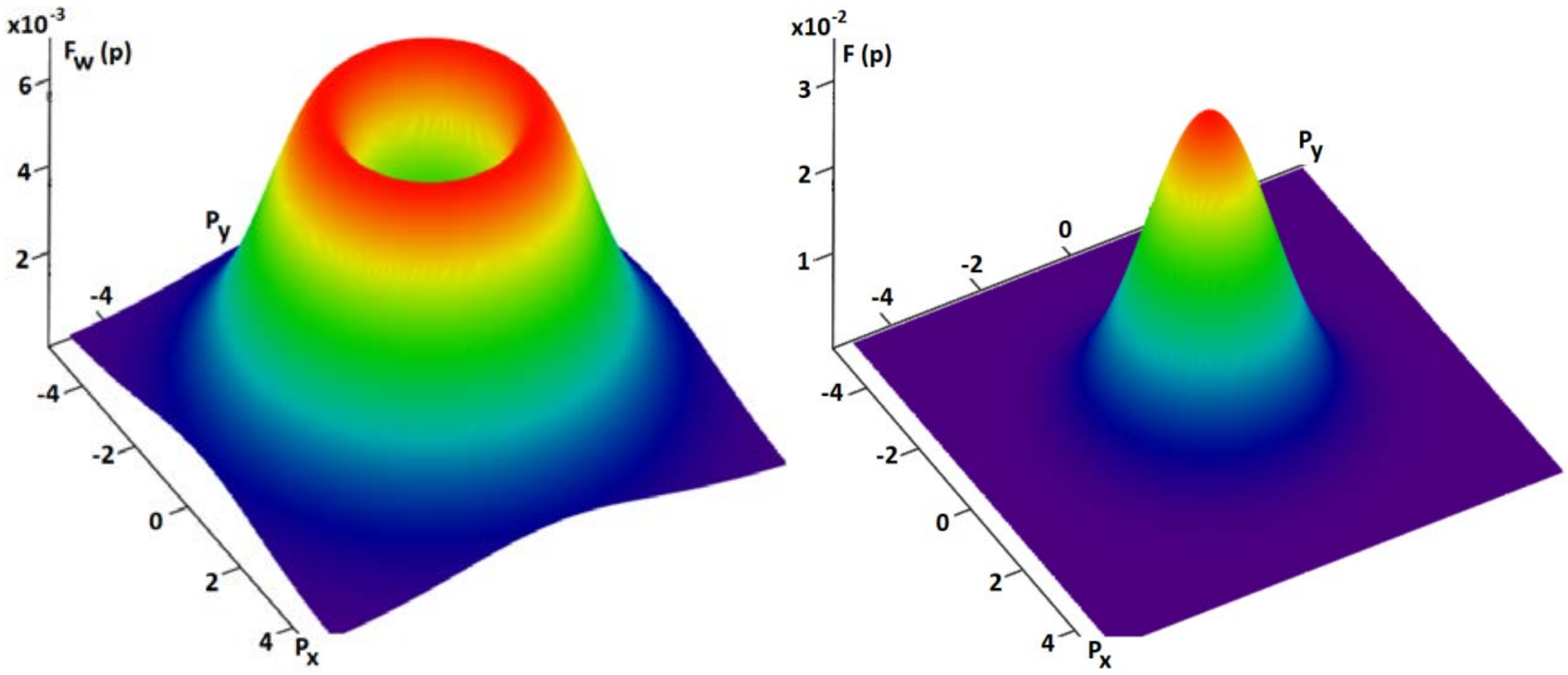

Fig. 4 Momentum distribution for the system with potential $\vec{A}_1$

The Wigner function (i.22) does not explicitly contain any dependency upon the vector potential $\vec{A}$, and it is only defined by the wave function. If the wave function $\Psi^{(EM)}$ is the same then function $W$ is also the same for two systems. The density function $f_w$ (i.27) depends explicitly upon the vector potential $\vec{A}$. Consequently, the expressions for density distribution (1.39)-(1.40) differ from each other. Fig. 4 shows the momentum distribution density (1.39) and $\left|\tilde{\Psi}^{(EM)}\right|^2$ on the $P_z = 0$ plane. On the left of Fig. 4 the distribution (1.39) is shown of the system with vector potential $\vec{A}_1$, and distribution $\left|\tilde{\Psi}^{(EM)}\right|^2$ is on the right. Comparing these distributions in Fig. 4 we can see their essential difference.

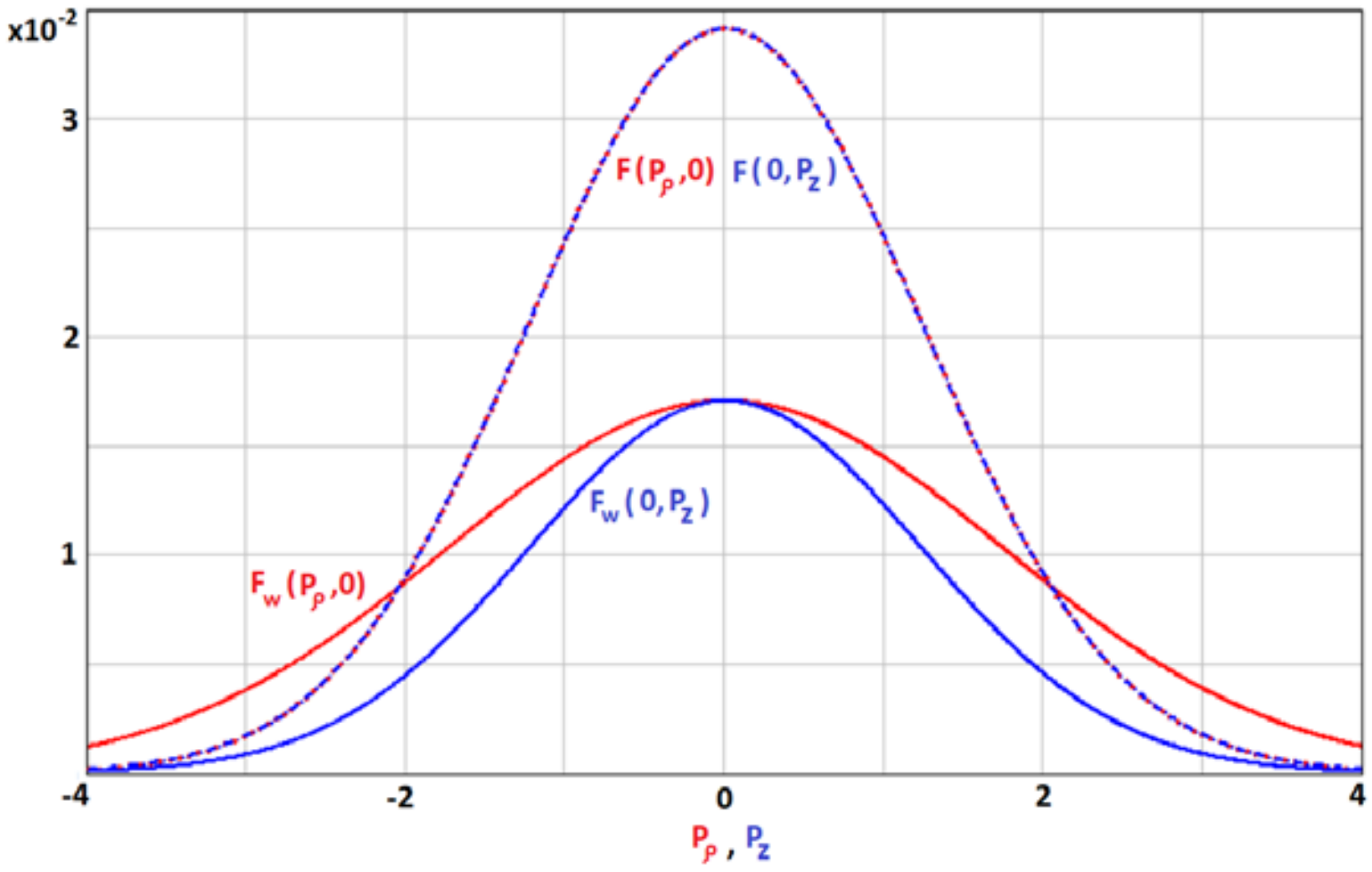

Fig. 5 Momentum distribution for the system with potential $\vec{A}_2$

Fig. 5 illustrates distributions along the axial axis $P_z$ and radial axis $P_\rho$. The red curve stands for the radial ($\rho$) distribution, and the blue curve corresponds to the longitudinal ($z$) distribution. The distribution (1.40) for vector potential $\vec{A}_2$ is represented by the solid line, and distribution $\left|\tilde{\Psi}^{(EM)}\right|^2$ is shown with the dash line. The distribution (1.41) is symmetric along $P_\rho$ and $P_z$, therefore, the red and the blue dash lines totally coincide in Fig. 5. The distribution (1.40) is not symmetric, so the red and the blue dash lines differ in Fig. 5. If we compare the distribution (1.40) and (1.41) in Fig. 5 then their essential difference will clearly be seen, which conforms with the inequality (1.41).

Let us note the distribution $\left|\tilde{\Psi}^{(EM)}\right|^2$ corresponds to the momentum operator $\hat{\mathrm{p}}$ (i.11), and the distributions (1.39) and (1.40) are constructed by function $f_w$. The following question may arise: «Which distribution can be physically reasonable»? The Wigner function $W$ (i.22) was phenomenologically obtained but meeting the relationships (1.37). The function $f_w$ (i.27) is derived on the $W$ basis, but it meets the conditions (1.38), (1.41). Is it possible that the electromagnetic momentum operator $\hat{\mathrm{P}} = \hat{\mathrm{p}} - q\vec{A}(\vec{r},t)$ possesses eigenfunctions $\psi : \hat{\mathrm{P}}\psi = \vec{\lambda}\psi$, by which the wave function $\Psi^{(EM)}$ should be decomposed? As a result, can we obtain $\tilde{\tilde{\Psi}}^{(EM)}(\vec{P},t)$ meeting $\left|\tilde{\tilde{\Psi}}^{(EM)}\right|^2 = \int f_w d^3r$? Let us construct functions $\psi$ and obtain

$$\hat{\mathrm{P}}\psi = -i\hbar\nabla_r\psi - q\vec{A}(\vec{r},t)\psi = \vec{\lambda}\psi,$$
$$\vec{\lambda} = -i\hbar\nabla_r \operatorname{Ln}\psi - q\vec{A}(\vec{r},t), \tag{1.42}$$

where $\vec{\lambda}$ does not depend on $\vec{r}$. Expression (1.42) coincides with the Helmholtz decomposition (i.12) under condition (i.14). Field $q\vec{A}(\vec{r},t)$ is vortex, and $-i\hbar\nabla_r \operatorname{Ln}\psi$ is potential one. Vortex and potential fields cannot compensate each other and produce field $\vec{\lambda}$ independent of $\vec{r}$. This «compensation» is possible in special cases for non-smooth potentials, for example, (1.25).

From comparison of (1.42), (i.12) and (i.14), the solution follows

$$\vec{\lambda} = m\langle \vec{v} \rangle = \langle \vec{p} \rangle(\vec{r},t), \ \psi(\vec{r},t) = \exp\left[ i\varphi(\vec{r},t) \right]. \tag{1.43}$$

In the special case at $\vec{A} = \vec{\theta}$, phase (action) $\hbar\varphi$ may be selected as:

$$\hbar\varphi(\vec{r},t) = \vec{r} \cdot \vec{p} - \mathrm{E}t,$$

that is the eigenfunctions (1.43) will correspond to the eigenfunctions $\psi = \exp\left( i\dfrac{\vec{p} \cdot \vec{r}}{\hbar} \right)$ of momentum operator $\hat{\mathrm{p}}$.

## §2 Motion equations

In papers [29-37], analogues of the Moyal equations for the Wigner functions (i.22) and (i.27) were obtained taking into account the electromagnetic field. The resulting equations for the evolution of the Wigner function have a form that does not coincide with the form of the second Vlasov equation (i.3). In order to construct the Vlasov-Moyal approximation taking the electromagnetic field for $\langle \dot{\vec{v}} \rangle$ into account it is necessary that the evolution equation of the Wigner function coincide with the second Vlasov equation.

Based on expressions (1.30), (1.33), (1.34) and (1.35), the use of the Wigner function (i.22) in the analysis of systems with electromagnetic interaction makes sense. Also, the expression for function (i.22) is much simpler than expression (i.27).

Let us construct an evolution equation for the function (i.22), reducible to the second Vlasov equation (i.3).

**Theorem 5** *Wigner function $W(\vec{r},\vec{p},t)$ of a pure state for quantum systems with electromagnetic interaction (i.13) satisfies the evolutionary equation:*

$$\frac{\partial}{\partial t}W + \frac{1}{m}\left( \vec{p} - q\overrightarrow{\hat{\mathcal{A}}} \right) \cdot \nabla_r W - \left( \nabla_p W \cdot \nabla_r \right)\left[ \frac{1}{2m}\left( \vec{p} - q\overleftarrow{\hat{\mathcal{A}}} \right)^2 + \overleftarrow{\hat{\mathcal{U}}} \right] = 0, \tag{2.1}$$

*or*

$$\frac{\partial}{\partial t}W + \frac{1}{m}\mathrm{div}_r\left[ W\overleftarrow{\hat{\mathcal{P}}} \right] + \mathrm{div}_p\left[ -W\vec{\nabla}_r\left( \frac{1}{2m}\overleftarrow{\hat{\mathcal{P}}}^2 + \overleftarrow{\hat{\mathcal{U}}} \right) \right] = 0, \tag{2.2}$$

*where*

$$\overrightarrow{\hat{\mathcal{P}}} \stackrel{\text{det}}{=} m\overrightarrow{\hat{\mathcal{V}}} \stackrel{\text{det}}{=} \vec{p} - q\overrightarrow{\hat{\mathcal{A}}}, \tag{2.3}$$

$$\overleftarrow{\hat{\mathcal{U}}} \stackrel{\text{det}}{=} \sum_{l=0}^{+\infty} \frac{(-1)^l (\hbar/2)^{2l}}{(2l+1)!} \left( \overleftarrow{\nabla}_p \cdot \overrightarrow{\nabla}_r \right)^{2l} U, \qquad \overrightarrow{\hat{\mathcal{A}}} \stackrel{\text{det}}{=} \sum_{k=0}^{+\infty} \frac{(-1)^k (\hbar/2)^{2k}}{(2k)!} \vec{A}\left( \overleftarrow{\nabla}_r \cdot \overrightarrow{\nabla}_p \right)^{2k}, \tag{2.4}$$

$$\overrightarrow{\hat{\mathcal{U}}} \stackrel{\text{det}}{=} \sum_{l=0}^{+\infty} \frac{(-1)^l (\hbar/2)^{2l}}{(2l+1)!} U\left( \overleftarrow{\nabla}_r \cdot \overrightarrow{\nabla}_p \right)^{2l}, \qquad \overleftarrow{\hat{\mathcal{A}}} \stackrel{\text{det}}{=} \sum_{k=0}^{+\infty} \frac{(-1)^k (\hbar/2)^{2k}}{(2k)!} \left( \overleftarrow{\nabla}_p \cdot \overrightarrow{\nabla}_r \right)^{2k} \vec{A}, \tag{2.5}$$

*and the arrow above the operator shows the direction of its action.*

The proof of Theorem 5 is given in Appendix B.

**Remark**

Operators (2.4) and (2.5) correspond to the scalar and vector potentials of the electromagnetic field. In the «classical limit» for $\hbar \to 0$, series (2.4) and (2.5) contain only one summand, the first summand, which coincides with the classical potential ($U/q$ or $\vec{A}$). A similar situation occurs when the potentials $U/q$ and $\vec{A}$ are linearly dependent on coordinate $\vec{r}$ (uniform electromagnetic field). Even for $\hbar \neq 0$ all derivatives $(\nabla_r)^{2k}$, $k>0$ in series (2.4)-(2.5) is equal to zero.

In the general case, when $\hbar \neq 0$ and the potentials are nonlinear, there are other terms of the series containing factor $\hbar^{2k}$ and formally corresponding to the «quantum» contributions [39].

The expression on the right (the last/third summand) in equation (2.2) can be considered as the electromagnetic Hamiltonian operator $\hat{\vec{\mathcal{H}}} = \frac{1}{2m}\hat{\vec{\mathcal{P}}}^2 + \hat{\vec{\mathcal{U}}}$, leading to the operator analogue of the Hamilton-Jacobi equation (i.15).

Equation (2.2) has a form similar to the second Vlasov equation (i.3), but by virtue of Theorem 4, it is convenient to go from momentum $\vec{p}$ to momentum $\vec{P} = \vec{p} - q\vec{A}$.

**Theorem 6** *Equation (2.1) for the Vlasov function* $f_V(\vec{r},\vec{P},t) = m^3 W(\vec{r},\vec{p},t)$ *has the form:*

$$\frac{\partial f_V}{\partial t} + \frac{1}{m}\hat{\vec{\mathcal{P}}}\cdot\nabla_r f_V + \hat{\vec{\mathcal{F}}}\cdot\nabla_P f_V = 0, \qquad (2.6)$$

*or*

$$\frac{\partial f_V}{\partial t} + \operatorname{div}_r\left(\hat{\vec{\mathcal{V}}} f_V\right) + \operatorname{div}_P\left[q f_V\left(\hat{\vec{\mathcal{E}}} + \hat{\vec{\mathcal{V}}}\times\hat{\vec{\mathcal{B}}} + \frac{\vec{\mathrm{d}}}{\mathrm{dt}}\hat{\vec{\mathcal{A}}}^{(\hbar)}\right)\right] = 0, \qquad (2.7)$$

*where*

$$\hat{\vec{\mathcal{F}}} = \hat{\vec{\mathcal{F}}}^{(EM)} + \hat{\vec{\mathcal{F}}}^{(\hbar)}, \qquad \hat{\vec{\mathcal{F}}}^{(EM)} \overset{\det}{=} q\left(\hat{\vec{\mathcal{E}}} + \frac{1}{m}\hat{\vec{\mathcal{P}}}\times\hat{\vec{\mathcal{B}}}\right), \qquad \hat{\vec{\mathcal{F}}}^{(\hbar)} \overset{\det}{=} q\frac{\vec{\mathrm{d}}}{\mathrm{dt}}\hat{\vec{\mathcal{A}}}^{(\hbar)}, \qquad (2.8)$$

$$\hat{\vec{\mathcal{A}}}^{(\hbar)} \overset{\det}{=} \sum_{k=1}^{+\infty}\frac{(-1)^k(\hbar/2)^{2k}}{(2k)!}\vec{A}\left(\overleftarrow{\nabla}_r\cdot\vec{\nabla}_P\right)^{2k},$$

$$\frac{\vec{\mathrm{d}}}{\mathrm{dt}} \overset{\det}{=} \frac{\partial}{\partial t} + \hat{\vec{\mathcal{V}}}\cdot\vec{\nabla}_r, \qquad \hat{\vec{\mathcal{B}}} \overset{\det}{=} \operatorname{curl}_r\hat{\vec{\mathcal{A}}}, \qquad \hat{\vec{\mathcal{E}}} \overset{\det}{=} -\frac{\partial}{\partial t}\hat{\vec{\mathcal{A}}} - \frac{1}{q}\nabla_r\hat{\vec{\mathcal{U}}}, \qquad (2.9)$$

*with that a substitution of* $\nabla_p$ *for* $\nabla_P$ *is made in operator* $\hat{\vec{\mathcal{A}}}$ *(2.4).*

The proof of Theorem 6 is given in Appendix B.

By comparing the second Vlasov equation (i.3) and equation (2.7), it is possible to extend the Vlasov-Moyal approximation (i.24) for the electromagnetic system.

**Definition** *We will call the approximation of the average acceleration flow field in the phase space of the form:*

$$f_V\left\langle \dot{\vec{P}} \right\rangle = qf_V\left( \overleftarrow{\hat{\vec{\mathcal{E}}}} + \overleftarrow{\hat{\vec{\mathcal{V}}}}\times\overleftarrow{\hat{\vec{\mathcal{B}}}} + \frac{\overleftarrow{\mathrm{d}}}{\mathrm{dt}}\overleftarrow{\hat{\vec{\mathcal{A}}}}^{(\hbar)} \right), \tag{2.10}$$

*the Vlasov-Moyal approximation for a quantum system with an electromagnetic field.*

The contributions of $qf_V\overleftarrow{\hat{\vec{\mathcal{E}}}}$ and $qf_V\overleftarrow{\hat{\vec{\mathcal{V}}}}\times\overleftarrow{\hat{\vec{\mathcal{B}}}}$ are infinite series (2.9), (2.4)-(2.5). The first term in the series $qf_V\overleftarrow{\hat{\vec{\mathcal{E}}}}$ is the classical Coulomb force, and, in $qf_V\overleftarrow{\hat{\vec{\mathcal{V}}}}\times\overleftarrow{\hat{\vec{\mathcal{B}}}}$, it is the classical Lorentz force. The remaining terms of the series have coefficients $\hbar^{2k}$ and can be interpreted as quantum corrections to the Coulomb and Lorentz forces. The third term $qf_V\frac{\overleftarrow{\mathrm{d}}}{\mathrm{dt}}\overleftarrow{\hat{\vec{\mathcal{A}}}}^{(\hbar)}$ is associated with the presence of a magnetic field and is of exclusively quantum nature, since it contains only coefficients $\hbar^{2k}$ at $k\in\mathbb{N}$.

In the absence of a magnetic field ($\vec{A}=\vec{\theta}$), approximation (2.10) transforms into approximation (i.24). Indeed, taking (2.4) and (2.9) into account, we obtain

$$f_V\left\langle \dot{\vec{P}} \right\rangle = qf_V\overleftarrow{\hat{\vec{\mathcal{E}}}} = -f_V\nabla_r\overleftarrow{\hat{\mathcal{U}}} = -\sum_{l=0}^{+\infty}\frac{(-1)^l(\hbar/2)^{2l}}{(2l+1)!}f_V\left(\overleftarrow{\nabla}_P\cdot\vec{\nabla}_r\right)^{2l}\nabla_r U. \tag{2.11}$$

**Theorem 7** *Averaging the Vlasov-Moyal approximation (2.10) over momentum space $\vec{P}$ gives an external force in hydrodynamic equation of motion (i.11) of the form:*

$$\left\langle\left\langle \dot{\vec{P}} \right\rangle\right\rangle = q\left( -\frac{\partial\vec{A}}{\partial t} - \frac{1}{q}\nabla_r U \right) + \frac{q}{m}\left\langle\vec{P}\right\rangle\times\vec{B}, \tag{2.12}$$

$$\int_{\mathbb{R}^3} f_V\overleftarrow{\hat{\vec{\mathcal{P}}}}d^3P = f_1\left\langle\vec{P}\right\rangle. \tag{2.13}$$

The proof of Theorem 7 is given in Appendix B.

**Remark**

From expression (2.13) it follows that the averaging of equation (2.7) over the momentum transforms it into the first Vlasov equation (i.2).

$$\frac{\partial f_1}{\partial t} + \frac{1}{m}\mathrm{div}_r\left( f_1\left\langle\vec{P}\right\rangle\right) = 0. \tag{2.14}$$

Quantum corrections (terms in series (2.4)-(2.5)) with coefficients $\hbar^{2k}$ when averaging over momentum $P$ completely disappeared from expression (2.12), transforming it into the classical expression for the external electromagnetic force.

Note that equation (2.7) itself is similar to the second Vlasov equation (i.3), but not equivalent to it. The difference between the equations is contained in the second term $\mathrm{div}_r\left( f_V\overleftarrow{\hat{\vec{\mathcal{P}}}}\right) \neq \mathrm{div}_r\left( f_2\vec{P}\right)$. Only the first term in the expression

$$\mathrm{div}_r\left( f_V\overleftarrow{\hat{\vec{\mathcal{P}}}}\right) = \mathrm{div}_r\left( f_V\vec{P}\right) - q\,\mathrm{div}_r\left( f_V\overleftarrow{\hat{\vec{\mathcal{A}}}}^{(\hbar)}\right), \tag{2.15}$$

coincides with the second term in the Vlasov equation (i.3). The difference is made by quantum corrections $f_V \overleftarrow{\mathcal{A}}^{(\hbar)}$. However, approximation (2.10) can be used to cut the chain of Vlasov equations off on the second equation. In this case, the second Vlasov equation (i.3) for the function $f_2\left(\vec{r},\vec{P},t\right)$ takes the form:

$$\frac{\partial f_2}{\partial t} + \frac{1}{m}\mathrm{div}_r\left(f_2\vec{P}\right) + \mathrm{div}_P\left[f_2\left\langle\dot{\vec{P}}\right\rangle\right] = 0, \tag{2.16}$$

The equation (2.16), integrated over momentum space $\vec{P}$, like equation (2.7), will transform into the first Vlasov equation (2.14).

Let us compare hydrodynamic equation (i.11) with electromagnetic equation (i.16) and the averaged approximation (2.12). Thus, we obtain the relation

$$\frac{1}{m}\frac{\partial \mathrm{Q}}{\partial x^k} = \frac{1}{f_1}\frac{\partial \mathrm{P}_{k\lambda}}{\partial x^\lambda},$$

which completely coincides with the previously obtained expression (i.26) in the absence of a magnetic field.

*Therefore, the presence of a magnetic field does not affect the quantum pressure.* Indeed, exact solutions (1.8) and (1.9) have the same quantum potential (1.14), although one system has a magnetic field and the other does not.

Despite the fact that the external force (2.12) in equation of motion (i.11) is classical, the equation itself describes the dynamics of a quantum system, since it contains quantum pressure (i.15)/(i.26). Quantum pressure $\nabla_r \mathrm{Q}$ arises as a counteraction to external force $-\nabla_r U$ that holds the particle in potential $U$. In a sense, a quantum system is a macroscopic object (in phase space), for which a hydrodynamic description is applicable. The absence of external potential $U$ will lead to the disappearance of quantum pressure as well as to dispersion of the wave packet. This analogy is similar to the behavior of a gas that is released from a vessel.

Note that approximation (2.10) is not the only one possible for the average acceleration flow $\left\langle\dot{\vec{P}}\right\rangle$. The problem is that evolution equation (2.7) contains information only on $\mathrm{div}_P\left[f_V\left\langle\dot{\vec{P}}\right\rangle\right]$, and according to the Helmholtz theorem, to correctly restore the field $\left\langle\dot{\vec{P}}\right\rangle$, the information about vortex component $\mathrm{curl}_P$ is needed, which is missing in equation (2.7). From a physical point of view, uncertainty arises in the construction of analogues of the «trajectories» of a quantum system according to approximation (2.10).

On the one hand, such uncertainty in trajectories can be attributed to the Heisenberg uncertainty principle. Indeed, during the classical transition $\hbar \to 0$, the terms with coefficients $\hbar^{2k}$ in the scalar $\hat{\mathcal{U}}$ and vector $\hat{\mathcal{A}}$ potential operators (2.4)-(2.5) will disappear. Consequently, approximation (2.10) transforms into

$$\lim_{\hbar\to 0}\left\langle\dot{\vec{P}}\right\rangle = q\left(-\frac{\partial\vec{A}}{\partial t} - \frac{1}{q}\nabla_r U\right) + \frac{q}{m}\vec{P}\times\vec{B}. \tag{2.17}$$

Expression (2.17) does not contain quantum corrections, and quantum potential $\mathrm{Q}\to 0$ at $\hbar \to 0$, that is, equations of motion (i.11) and (i.16) will determine the classical trajectories of motion without quantum pressure. Note that expression (2.17) was phenomenologically introduced by Vlasov when the chain on the second equation was cut off for the solving of the plasma physics problems.

On the other hand, as shown in [25], the transition from quantum to classical systems is not necessarily associated with the limit $\hbar \to 0$. When directly substituting the Wigner function into the Vlasov-Moyal approximation, Planck's constant $\hbar$ is not present explicitly. As a result, the uncertainty in the trajectory is associated with the spatial scale $\sigma_r$.

## Conclusion

An important difference between function $f_w$ and the Wigner function $W$ is that it has negative values for wave function $\Psi_G \sim \exp\left(-r^2/2\sigma_r^2\right)$. Note that, according to the Hudson's theorem [22] and its generalization to the 3D case [23], the Wigner function $W$ is positive only for $\Psi_G$. Thus, function $f_w$ does not satisfy this condition.

It is shown that function $f_w$ is positive (1.27) in the entire phase region for an electromagnetic quantum system (1.26) described by a non-Gaussian wave function (1.8). In this case, the opposite sign of vector potential $\vec{A}_3$ (1.22) and scalar potential $\Phi \sim \phi$ is essential. In fact, Corollary 2 is an analogue of the Hudson theorem and its extension to the 3D case for function $f_w$.

Wave function $\Psi_G$ is a solution to the Schrödinger equation for a harmonic oscillator with potential $U_0 \sim x^2$. Substituting potential $U_0$ into the right-hand side of equation (i.23), we obtain zero. As a result, the Moyal equation becomes the classical Liouville equation with positive solution $W_G$. In the example considered in this paper, potential $U_1$ (1.12) has the form of a quadratic funnel with a pole at the origin of coordinates. Substitution of potential (1.12) in series (2.4)-(2.5) leads to the presence of an infinite number of nonzero summands. Thus, equation (2.1) differs from the Liouville equation. Despite this fact, solution $W_G$ of equation (2.1) is positive. This behavior is associated with the presence of a magnetic field. In the absence of a magnetic field, the wave function for the potential (1.12) differs from the Gaussian distribution (1.8) and the Wigner function has negative values.

An analysis of the properties of probability quasi-density functions $W$ and $f_w$ on the model example (1.8)-(1.9) shows the correctness of calculating the mean values over the phase space (lemma 2) (1.30)-(1.36) for an electromagnetic system. The expression for function $W$ (i.22) is much simpler than for function $f_w$ (i.27). A similar statement is also true for the evolutionare equations satisfied by functions $W$ and $f_w$.

An important difference between functions $W$ and $f_w$ is the different type of momentum distribution (Corollary 3) (1.39)-(1.41).

Function $f_V\left(\vec{r},\vec{P},t\right) = m^3 W\left(\vec{r},\vec{p},t\right)$ for electromagnetic systems satisfies equation (2.7). A significant difference from the known forms of equation (2.7) is its similarity with the second Vlasov equation (i.3). This fact allows us to expand the Vlasov-Moyal approximation (i.24) to systems with electromagnetic interaction in the form (2.10). The first term in approximation (2.10) has the form (2.17) and coincides with the well-known Vlasov approximation (i.20)-(i.21), used in plasma physics, astrophysics, and solid state physics. This term does not contain quantum corrections, since it has coefficient $\hbar^0$. The following terms in approximation (2.10) according to (2.4)-(2.5) contain coefficients $\hbar^{2k}$ and give quantum corrections to the external force $\left\langle \dot{\vec{P}} \right\rangle$ in the second Vlasov equation (2.16). Using expression (2.10) one can find the average sources of dissipations $\langle\langle Q_2\rangle\rangle=\left\langle\left\langle \mathrm{div}_P\left\langle \dot{\vec{P}}\right\rangle\right\rangle\right\rangle$ for evolution equation (i.9) of the Boltzmann $H_2$–function. An interesting fact is the disappearance of quantum contributions when averaging $\left\langle\left\langle \dot{\vec{P}}\right\rangle\right\rangle$ over momentum space $\vec{P}$. Here the external force $\left\langle\left\langle \dot{\vec{P}}\right\rangle\right\rangle$ contains only classical terms (2.12). In this case, in equation of motion (i.11), the quantum information remains only in the form of quantum pressure $\nabla_r \mathrm{Q}$. As it turned out from the obtained approximation (2.10), the quantum pressure itself is not explicitly related to the presence of a magnetic field. The physical meaning of quantum term $qf_V \frac{\overleftarrow{\mathrm{d}}}{\mathrm{dt}}\bar{\hat{\mathcal{A}}}^{(\hbar)}$ associated with the magnetic field remains an open question.

## Acknowledgements

This research has been supported by the Interdisciplinary Scientific and Educational School of Moscow University «Photonic and Quantum Technologies. Digital Medicine».

## Appendix A

### *Proof of Theorem 1*

To prove the theorem, one can use the method described in [4, 7] or make a direct substitution. Without loss of generality, we use the second method. For the electromagnetic case (1.9), (1.11), we obtain:

$$U=E+\frac{\hbar^2}{2m}\frac{\Delta_r|\Psi|}{|\Psi|}-\frac{q^2}{2m}\left|\vec{A}\right|^2-i\hbar\frac{q}{m|\Psi|}\vec{A}\cdot\nabla_r|\Psi|, \tag{A.1}$$

$$U_1=E-\mathrm{Q}-\frac{\hbar^2}{2m\rho^2}, \tag{A.2}$$

where it is taken into account that $\mathrm{Q}=-\frac{\hbar^2}{2m}\frac{\Delta_r|\Psi|}{|\Psi|}$, $\vec{A}_1=-\frac{\hbar}{q\rho}\vec{e}_\phi$, $\mathrm{div}_r\vec{A}_1=0$ and $\varphi=-\frac{E}{\hbar}t$. The expression for the quantum potential has the form:

$$\Delta_r|\Psi|=\frac{N}{2\sigma_r^2}e^{-\frac{\rho^2}{4\sigma_r^2}-\frac{z^2}{4\sigma_r^2}}\left(\frac{\rho^2}{2\sigma_r^2}-2\right)+\frac{N}{2\sigma_r^2}e^{-\frac{\rho^2}{4\sigma_r^2}-\frac{z^2}{4\sigma_r^2}}\left(\frac{z^2}{2\sigma_r^2}-1\right),$$

$$\mathrm{Q}=-\frac{\hbar^2}{4m\sigma_r^2}\left(\frac{\rho^2}{2\sigma_r^2}-3+\frac{z^2}{2\sigma_r^2}\right), \tag{A.3}$$

where $N^{-1}=(2\pi)^{3/4}\sigma_r^{3/2}$. For the case with electrical interaction only, the proof is given in [25]. Theorem 1 is proved.

### *Proof of Theorem 2*

Using expression (A.1), we obtain

$$U_2 = E - \mathrm{Q} - \frac{\hbar^2\eta^2\rho^2}{8m\sigma_r^4} + \frac{i\hbar^2\eta\rho}{2\sigma_r^2 m}\vec{e}_\phi \cdot \left( -\frac{\rho}{2\sigma_r^2}\vec{e}_\rho - \frac{z}{2\sigma_r^2}\vec{e}_z \right). \tag{A.4}$$

Since the form of the wave function has not changed, the quantum potential (A.3) has not changed either. Substituting (A.3) into (A.4), we derive (1.17). Theorem 2 is proved.

***Proof of Theorem 3***

Using the Helmholtz decomposition for the vector field $\langle \vec{p} \rangle = m\langle \vec{v} \rangle$, we get:

$$\langle \vec{p} \rangle = m\langle \vec{v} \rangle = -\alpha m \nabla_r \Phi = -\frac{i\hbar}{2}\nabla_r \operatorname{Ln}\left(\frac{\Psi}{\bar{\Psi}}\right) = -\frac{i\hbar}{2}\left(\frac{\nabla_r \Psi}{\Psi} - \frac{\nabla_r \bar{\Psi}}{\bar{\Psi}}\right),$$

$$|\Psi|^2 \langle \vec{p} \rangle = -\frac{i\hbar}{2}\left(\bar{\Psi}\nabla_r \Psi - \Psi \nabla_r \bar{\Psi}\right) = \int_{\mathbb{R}^3} f_2(\vec{r}, \vec{p}, t)\, \vec{p} d^3 p. \tag{A.5}$$

Let us transform expression (A.5), bringing the left side of (A.5) to an integral form. Let us make a change of variables $\vec{r}_\pm = \vec{r} \pm \vec{s}/2$ and note that

$$\nabla_s \left[\bar{\Psi}(\vec{r}_-, t)\Psi(\vec{r}_+, t)\right] = \frac{1}{2}\left[\bar{\Psi}(\vec{r}_-, t)\nabla_{r_+}\Psi(\vec{r}_+, t) - \Psi(\vec{r}_+, t)\nabla_{r_-}\bar{\Psi}(\vec{r}_-, t)\right]. \tag{A.6}$$

Expression (A.6) will transform into expression (A.5) at $s = 0$, therefore,

$$|\Psi|^2 \langle \vec{p} \rangle = -i\hbar \int_{\mathbb{R}^3} \delta(\vec{s}) \nabla_s \left[\Psi(\vec{r}_+, t)\bar{\Psi}(\vec{r}_-, t)\right] d^3 s =$$

$$= \frac{-i\hbar}{(2\pi\hbar)^3} \int_{\mathbb{R}^3} e^{-i\frac{\vec{p}\cdot\vec{s}}{\hbar}} d^3 p \int_{\mathbb{R}^3} \nabla_s \left[\Psi(\vec{r}_+, t)\bar{\Psi}(\vec{r}_-, t)\right] d^3 s =$$

$$= \frac{-i\hbar}{(2\pi\hbar)^3} \int_{\mathbb{R}^3} d^3 p \left\{ e^{-i\frac{\vec{p}\cdot\vec{s}}{\hbar}} \Psi(\vec{r}_+, t)\bar{\Psi}(\vec{r}_-, t) \Big|_{\pm\infty} - \int_{\mathbb{R}^3} \Psi(\vec{r}_+, t)\bar{\Psi}(\vec{r}_-, t) \nabla_s e^{-i\frac{\vec{p}\cdot\vec{s}}{\hbar}} d^3 s \right\},$$

$$|\Psi|^2 \langle \vec{p} \rangle = \frac{1}{(2\pi\hbar)^3} \int_{\mathbb{R}^3} \vec{p} \left[ \int_{\mathbb{R}^3} \Psi(\vec{r}_+, t)\bar{\Psi}(\vec{r}_-, t) e^{-i\frac{\vec{p}\cdot\vec{s}}{\hbar}} d^3 s \right] d^3 p. \tag{A.7}$$

Comparing expressions (A.5) and (A.7), we get (1.19). Theorem 3 is proved.

***Proof of Lemma 1***

Substituting expression (1.9) into function (i.27), we obtain

$$(2\pi\hbar)^3 (2\pi)^{3/2} \sigma_r^3 f_w(\vec{r}, \vec{P}) = \int_{\mathbb{R}^3} e^{-\frac{1}{4\sigma_r^2}\left[\left(\vec{r}+\frac{\vec{s}}{2}\right)^2 + \left(\vec{r}-\frac{\vec{s}}{2}\right)^2\right]} e^{-\frac{i}{\hbar}\vec{s}\cdot\vec{P}'} d^3 s = e^{-\frac{r^2}{2\sigma_r^2}} \int_{\mathbb{R}^3} e^{-\frac{s^2}{8\sigma_r^2} - \frac{i}{\hbar}\vec{s}\cdot\vec{P}'} d^3 s, \tag{A.8}$$

where $\vec{P}' = \vec{P} + \frac{q}{2}\int_{-1}^{1} \vec{A}\left(\vec{r} + \tau\frac{\vec{s}}{2}\right) d\tau$. Let us calculate the integral $\frac{\vec{s}}{2} q \int_{-1}^{1} \vec{A}\left(\vec{r} + \tau\frac{\vec{s}}{2}\right) d\tau$ with the field (1.12):

$$\frac{\vec{s}}{2}q\int_{-1}^{1}\vec{A}\left(\vec{r}+\tau\frac{\vec{s}}{2}\right)d\tau = q\int_{\vec{r}_-}^{\vec{r}_+}\vec{A}(\vec{r}')d\vec{r}' = -\hbar\int_{\vec{r}_-}^{\vec{r}_+}\frac{1}{\rho'}\vec{e}_{\phi'}d\vec{r}' = -\hbar\int_{\phi_-}^{\phi_+}d\phi' = -\hbar\left(\phi_+ - \phi_-\right), \qquad \text{(A.9)}$$

where $d\vec{r}' = \{d\rho', \rho' d\phi', dz'\}$, $\vec{r}_\pm = \vec{r} \pm \frac{\vec{s}}{2}$ and $\mathrm{tg}\,\phi_\pm = \frac{y \pm y_s/2}{x \pm x_s/2}$. When calculating the integral (A.9), the potential of field $\vec{A}$ outside pole $\rho = 0$ is taken into account. From (A.9), it follows that

$$\vec{s}\cdot\vec{P}' = \vec{s}\cdot\vec{P} - \hbar\left(\phi_+ - \phi_-\right). \qquad \text{(A.10)}$$

The lemma 1 is proved.

***Proof of Theorem 4***

Let us write integrand (i.22) of the Wigner function $W^{(\mathrm{E})}(\vec{r},\vec{p})$ for the wave function (1.8), we obtain [25]:

$$\bar{\Psi}^{(\mathrm{E})}(\vec{r}_-,t)\Psi^{(\mathrm{E})}(\vec{r}_+,t) = \frac{e^{i(\phi_+-\phi_-)}}{(2\pi)^{3/2}\sigma_r^3}e^{-\frac{1}{2\sigma_r^2}\left(r^2+\frac{s^2}{4}\right)}. \qquad \text{(A.11)}$$

In accordance with $\phi_\pm = \mathrm{arctg}\frac{y \pm y_s/2}{x \pm x_s/2}$, azimuth angles $\phi_\pm$ admit the representation:

$$\sin\phi_\pm = \frac{2y \pm y_s}{\sqrt{(2x\pm x_s)^2 + (2y\pm y_s)^2}} = \frac{\sin\phi \pm \rho_s'\sin\phi_s}{\sqrt{1+\rho_s'^2 \pm 2\rho_s'\cos(\phi-\phi_s)}}, \qquad \text{(A.12)}$$
$$\cos\phi_\pm = \frac{2x \pm x_s}{\sqrt{(2x\pm x_s)^2 + (2y\pm y_s)^2}} = \frac{\cos\phi \pm \rho_s'\cos\phi_s}{\sqrt{1+\rho_s'^2 \pm 2\rho_s'\cos(\phi-\phi_s)}},$$

where $\rho_s' = \frac{\rho_s}{2\rho}$. Substituting (A.11) into integral (i.22) gives expression (1.23) (with the change $\vec{p} \mapsto \vec{P}$) for the Wigner function $f_w(\vec{r},\vec{P})$. Theorem 4 is proved.

***Proof of Corollary 2***

Let us calculate expression (i.27) for function $f_w$, we obtain

$$f_w(\vec{r},\vec{P},t) = \frac{e^{-\frac{r^2}{2\sigma_r^2}}}{(2\pi\hbar)^3(2\pi)^{3/2}\sigma_r^3}\int_{\mathbb{R}^3} e^{-\frac{s^2}{8\sigma_r^2}-\frac{i}{\hbar}\vec{s}\cdot\vec{P}}d^3s = \frac{1}{(\pi\hbar)^3}e^{-\frac{r^2}{2\sigma_r^2}-\frac{2\sigma_r^2}{\hbar^2}P^2}, \qquad \text{(A.13)}$$

where expressions (1.8), (1.26), (A.9), (A.11) and (B.1) are taken into considerations. Let us check that the wave function (1.8) meets the Schrödinger equation (1.26). Here we use expression (A.1) for potential $U_1$ in which the sign of the vector potential (1.26) is not significant. The quantum potential results from expression (A.3).

Corollary 2 is proved.

***Proof of Lemma 2***

Taking expression (A.5) and the representation

$$\int_{\mathbb{R}^3} e^{-\frac{i}{\hbar}\vec{s}\cdot\vec{P}}\vec{P}d^3P = i\hbar\nabla_s \int_{\mathbb{R}^3} e^{-\frac{i}{\hbar}\vec{s}\cdot\vec{P}} d^3P = i\hbar\left(2\pi\hbar\right)^3 \nabla_s\delta\left(\vec{s}\right),$$

into account we come to

$$|\Psi|^2\left\langle\vec{P}\right\rangle = i\hbar\int_{\mathbb{R}^3} \exp\left(-\frac{i}{\hbar}q\int_{\vec{r}_-}^{\vec{r}_+}\vec{A}\left(\vec{r}'\right)d\vec{r}'\right)\rho\left(\vec{r}_+,\vec{r}_-,t\right)\nabla_s\delta\left(\vec{s}\right)d^3s = \tag{A.14}$$

$$= -i\hbar\int_{\mathbb{R}^3}\delta\left(\vec{s}\right)\nabla_s\left[\exp\left(-\frac{i}{\hbar}q\int_{\vec{r}_-}^{\vec{r}_+}\vec{A}\left(\vec{r}'\right)d\vec{r}'\right)\rho\left(\vec{r}_+,\vec{r}_-,t\right)\right]d^3s,$$

where integral $\int_{\vec{r}_-}^{\vec{r}_+}\vec{A}\left(\vec{r}'\right)d\vec{r}'$ is calculated along the straight line, connecting points $\vec{r}_\pm$. Let us transform the integrand (A.14)

$$\nabla_s\left[\exp\left(-\frac{i}{\hbar}q\int_{\vec{r}_-}^{\vec{r}_+}\vec{A}\left(\vec{r}',t\right)d\vec{r}'\right)\rho\left(\vec{r}_+,\vec{r}_-,t\right)\right] = \rho\left(\vec{r}_+,\vec{r}_-,t\right)\nabla_s e^{-\frac{i}{\hbar}q\int_{\vec{r}_-}^{\vec{r}_+}\vec{A}(\vec{r}',t)d\vec{r}'} + e^{-\frac{i}{\hbar}q\int_{\vec{r}_-}^{\vec{r}_+}\vec{A}(\vec{r}',t)d\vec{r}'}\nabla_s\rho\left(\vec{r}_+,\vec{r}_-,t\right) =$$

$$= -\frac{i}{2\hbar}q\rho\left(\vec{r}_+,\vec{r}_-,t\right)e^{-\frac{i}{\hbar}q\int_{\vec{r}_-}^{\vec{r}_+}\vec{A}(\vec{r}',t)d\vec{r}'}\nabla_s\int_{\vec{r}_-}^{\vec{r}_+}\vec{A}\left(\vec{r}',t\right)d\vec{r}' + \tag{A.15}$$

$$+\frac{1}{2}e^{-\frac{i}{\hbar}q\int_{\vec{r}_-}^{\vec{r}_+}\vec{A}(\vec{r}',t)d\vec{r}'}\left[\bar{\Psi}\left(\vec{r}_-,t\right)\nabla_{r_+}\Psi\left(\vec{r}_+,t\right) - \Psi\left(\vec{r}_+,t\right)\nabla_{r_-}\bar{\Psi}\left(\vec{r}_-,t\right)\right].$$

Substituting (A.15) into (A.14), we get

$$|\Psi|^2\left\langle\vec{P}\right\rangle = -\frac{\hbar}{2\hbar}q\left[\vec{A}\left(\vec{r},t\right)+\vec{A}\left(\vec{r},t\right)\right]\rho\left(\vec{r},\vec{r},t\right)e^{-\frac{i}{\hbar}q\int_{\vec{r}}^{\vec{r}}\vec{A}(\vec{r}',t)d\vec{r}'} -$$

$$-\frac{i\hbar}{2}e^{-\frac{i}{\hbar}q\int_{\vec{r}}^{\vec{r}}\vec{A}(\vec{r}',t)d\vec{r}'}\left[\bar{\Psi}\left(\vec{r},t\right)\nabla_r\Psi\left(\vec{r},t\right) - \Psi\left(\vec{r},t\right)\nabla_r\bar{\Psi}\left(\vec{r},t\right)\right],$$

$$|\Psi|^2\left\langle\vec{P}\right\rangle = \frac{i}{\hbar}q\vec{A}|\Psi|^2 - \frac{i\hbar}{2}\left[\bar{\Psi}\nabla_r\Psi - \Psi\nabla_r\bar{\Psi}\right] \tag{A.16}$$

or

$$\left\langle\vec{P}\right\rangle = -\frac{i\hbar}{2}\nabla_r \operatorname{Ln}\frac{\Psi}{\bar{\Psi}} - q\vec{A}. \tag{A.17}$$

Similarly for the function $W$, we obtain

$$|\Psi|^2\left\langle\vec{p}\right\rangle = \frac{1}{\left(2\pi\hbar\right)^3}\int_{\mathbb{R}^3} e^{-\frac{i}{\hbar}\vec{s}\cdot\vec{p}}\vec{p}d^3p\int_{\mathbb{R}^3}\rho\left(\vec{r}_+,\vec{r}_-,t\right)d^3s = -\frac{i\hbar}{2}\left[\bar{\Psi}\left(\vec{r},t\right)\nabla_r\Psi\left(\vec{r},t\right) - \Psi\left(\vec{r},t\right)\nabla_r\bar{\Psi}\left(\vec{r},t\right)\right],$$

$$\left\langle \vec{p} \right\rangle = -\frac{i\hbar}{2}\left[\frac{\nabla_r \Psi}{\Psi} - \frac{\nabla_r \bar{\Psi}}{\bar{\Psi}}\right], \quad \left\langle \vec{P} \right\rangle = \left\langle \vec{p} \right\rangle - q\vec{A}. \tag{A.18}$$

The lemma 2 is proved.

***Proof of Corollary 3***

The validity of condition (1.38) directly results from integration:

$$\begin{aligned} &\int_{\mathbb{R}^3} f_w\left(\vec{r},\vec{P},t\right) d^3P = \frac{1}{(2\pi\hbar)^3}\int_{\mathbb{R}^3} \bar{\Psi}\left(\vec{r}_-,t\right)\Psi\left(\vec{r}_+,t\right) e^{-\frac{iq}{2\hbar}\int_{-1}^{1}\vec{s}\cdot\vec{A}\left(\vec{r}+\vec{s}\frac{\tau}{2},t\right)d\tau} d^3s \int_{\mathbb{R}^3} e^{-\frac{i}{\hbar}\vec{P}\cdot\vec{s}} d^3P = \\ &= \int_{\mathbb{R}^3} \delta(s)\bar{\Psi}\left(\vec{r}_-,t\right)\Psi\left(\vec{r}_+,t\right) e^{-\frac{iq}{2\hbar}\int_{-1}^{1}\vec{s}\cdot\vec{A}\left(\vec{r}+\vec{s}\frac{\tau}{2},t\right)d\tau} d^3s = \left|\Psi\left(\vec{r},t\right)\right|^2. \end{aligned} \tag{A.19}$$

For the vector potential $\vec{A}_1$ expression (A.9) is valid, therefore

$$\begin{aligned} &\int_{\mathbb{R}^3} f_w\left(\vec{r},\vec{P}\right) d^3r = \frac{1}{(2\pi\hbar)^3}\int_{\mathbb{R}^3} \Psi^{(\mathrm{EM})}\left(\vec{r}_+,t\right) e^{i\phi_+} e^{-i\frac{\vec{P}\cdot\vec{r}_+}{\hbar}} d^3r_+ \int_{\mathbb{R}^3} \bar{\Psi}^{(\mathrm{EM})}\left(\vec{r}_-,t\right) e^{-i\phi_-} e^{i\frac{\vec{P}\cdot\vec{r}_-}{\hbar}} d^3r_- = \\ &= \frac{1}{(2\pi\hbar)^3}\int_{\mathbb{R}^3} \Psi^{(\mathrm{E})}\left(\vec{r}_+,t\right) e^{-i\frac{\vec{P}\cdot\vec{r}_+}{\hbar}} d^3r_+ \int_{\mathbb{R}^3} \bar{\Psi}^{(\mathrm{E})}\left(\vec{r}_-,t\right) e^{i\frac{\vec{P}\cdot\vec{r}_-}{\hbar}} d^3r_- = \tilde{\Psi}^{(\mathrm{E})}\left(\vec{P},t\right)\bar{\tilde{\Psi}}^{(\mathrm{E})}\left(\vec{P},t\right) = \left|\tilde{\Psi}^{(\mathrm{E})}\right|^2, \end{aligned} \tag{A.20}$$

where the gauge relation (1.24) is taken into account. Let us find $\tilde{\Psi}^{(\mathrm{E})}$

$$(2\pi\hbar)^{3/2}\,\tilde{\Psi}^{(\mathrm{E})}\left(\vec{p},t\right) = \int_{\mathbb{R}^3} \Psi^{(\mathrm{E})}\left(\vec{r},t\right) e^{-i\frac{\vec{p}\cdot\vec{r}}{\hbar}} d^3r = \int_{\mathbb{R}^3} \Psi^{(\mathrm{EM})}\left(\vec{r},t\right) e^{i\phi} e^{-i\frac{\vec{p}\cdot\vec{r}}{\hbar}} d^3r = \frac{e^{-i\frac{\mathrm{E}}{\hbar}t} I\left(\vec{p}\right)}{(2\pi)^{3/4}\sigma_r^{3/2}} \tag{A.21}$$

with

$$\begin{aligned} &I\left(\vec{p}\right) = \int_{\mathbb{R}^3} e^{-\frac{r^2}{4\sigma_r^2}} e^{i\phi} e^{-i\frac{\vec{p}\cdot\vec{r}}{\hbar}} d^3r = \\ &= 2\sigma_r\sqrt{\pi}\, e^{-\frac{\sigma_r^2}{\hbar^2}p_z^2} e^{i\phi_p} \int_{-\pi}^{\pi} e^{-\sigma_r^2\frac{p_\rho^2}{\hbar^2}\cos^2\phi'} d\phi' \int_0^{+\infty} e^{-\left[\frac{\rho}{2\sigma_r}+i\sigma_r\frac{p_\rho}{\hbar}\cos\phi'\right]^2} \rho\cos\phi' d\rho, \end{aligned} \tag{A.22}$$

where it is taken into account, that

$$\begin{aligned} &\int_0^{2\pi} e^{-i\frac{\rho p_\rho}{\hbar}\cos\left(\phi-\phi_p\right)} \sin\phi d\phi = \sin\phi_p \int_0^{2\pi} e^{-i\frac{\rho p_\rho}{\hbar}\cos\phi'} \cos\phi' d\phi', \\ &\int_0^{2\pi} e^{-i\frac{\rho p_\rho}{\hbar}\cos\left(\phi-\phi_p\right)} \cos\phi d\phi = \cos\phi_p \int_0^{2\pi} e^{-i\frac{\rho p_\rho}{\hbar}\cos\phi'} \cos\phi' d\phi'. \end{aligned}$$

We calculate the integral in expression (A.22), dividing it into two intervals $(-\pi,0)$ and $(0,\pi)$:

$$I_1 = \int_0^{\pi} e^{-\sigma_r^2 \frac{p_\rho^2}{\hbar^2}\cos^2\phi'} d\phi' \int_0^{+\infty} e^{-\left[\frac{\rho}{2\sigma_r} + i\sigma_r \frac{p_\rho}{\hbar}\cos\phi'\right]^2} \rho\cos\phi' d\rho = \int_0^{\pi/2} e^{-\sigma_r^2 \frac{p_\rho^2}{\hbar^2}\cos^2\phi'} d\phi' \int_0^{+\infty} e^{-\left[\frac{\rho}{2\sigma_r} + i\sigma_r \frac{p_\rho}{\hbar}\cos\phi'\right]^2} \rho\cos\phi' d\rho +$$
$$+ \int_{\pi/2}^{\pi} e^{-\sigma_r^2 \frac{p_\rho^2}{\hbar^2}\cos^2\phi'} d\phi' \int_0^{-\infty} e^{-\left[\frac{\rho'}{2\sigma_r} - i\sigma_r \frac{p_\rho}{\hbar}\cos\phi'\right]^2} \rho'\cos\phi' d\rho',$$

where $\rho = -\rho'$.

$$I_1 = \int_0^{\pi/2} e^{-\sigma_r^2 \frac{p_\rho^2}{\hbar^2}\cos^2\phi'} d\phi' \int_0^{+\infty} e^{-\left[\frac{\rho}{2\sigma_r} + i\sigma_r \frac{p_\rho}{\hbar}\cos\phi'\right]^2} \rho\cos\phi' d\rho + \int_{\pi/2}^{0} e^{-\sigma_r^2 \frac{p_\rho^2}{\hbar^2}\cos^2\phi''} d\phi' \int_0^{-\infty} e^{-\left[\frac{\rho'}{2\sigma_r} + i\sigma_r \frac{p_\rho}{\hbar}\cos\phi''\right]^2} \rho'\cos\phi'' d\rho' =$$
$$= \int_0^{\pi/2} e^{-\sigma_r^2 \frac{p_\rho^2}{\hbar^2}\cos^2\phi'} d\phi' \int_0^{+\infty} e^{-\left[\frac{\rho}{2\sigma_r} + i\sigma_r \frac{p_\rho}{\hbar}\cos\phi'\right]^2} \rho\cos\phi' d\rho + \int_0^{\pi/2} e^{-\sigma_r^2 \frac{p_\rho^2}{\hbar^2}\cos^2\phi''} d\phi'' \int_{-\infty}^{0} e^{-\left[\frac{\rho'}{2\sigma_r} + i\sigma_r \frac{p_\rho}{\hbar}\cos\phi''\right]^2} \rho'\cos\phi'' d\rho',$$
$$I_1 = \int_0^{\pi/2} e^{-\sigma_r^2 \frac{p_\rho^2}{\hbar^2}\cos^2\phi'} d\phi' \int_{-\infty}^{+\infty} e^{-\left[\frac{\rho}{2\sigma_r} + i\sigma_r \frac{p_\rho}{\hbar}\cos\phi'\right]^2} \rho\cos\phi' d\rho \qquad \text{(A.23)}$$

with $\phi'' = \pi - \phi'$. Similarly for integral $I_2$, we obtain

$$I_2 = \int_{-\pi}^{0} e^{-\sigma_r^2 \frac{p_\rho^2}{\hbar^2}\cos^2\phi'} d\phi' \int_0^{+\infty} e^{-\left[\frac{\rho}{2\sigma_r} + i\sigma_r \frac{p_\rho}{\hbar}\cos\phi'\right]^2} \rho\cos\phi' d\rho = \int_0^{\pi} e^{-\sigma_r^2 \frac{p_\rho^2}{\hbar^2}\cos^2\phi''} d\phi'' \int_0^{+\infty} e^{-\left[\frac{\rho}{2\sigma_r} + i\sigma_r \frac{p_\rho}{\hbar}\cos\phi''\right]^2} \rho\cos\phi'' d\rho, \qquad \text{(A.24)}$$

where $\phi'' = -\phi'$. Substituting (A.23) and (A.24) into (A.22), we arrive at the expression

$$I = -ie^{i\phi_p}\frac{16\pi\sigma_r^4}{\hbar} p_\rho e^{-\frac{\sigma_r^2}{\hbar^2}p_z^2} \int_0^{\pi/2} e^{-\sigma_r^2 \frac{p_\rho^2}{\hbar^2}\cos^2\phi'} \cos^2\phi' d\phi', \qquad \text{(A.25)}$$

where the property of the mean value of the Gaussian distribution is taken into account:

$$\int_{-\infty}^{+\infty} e^{-\left[\frac{\rho}{2\sigma_r} + i\sigma_r \frac{p_\rho}{\hbar}\cos\phi'\right]^2} \rho d\rho = -i4\sqrt{\pi}\sigma_r^3 \frac{p_\rho}{\hbar}\cos\phi'.$$

From expressions (A.25) and (A.21), we get

$$\tilde{\Psi}^{(\mathrm{E})}(p,t) = -ie^{i\phi_p}\frac{16\pi\sigma_r^4}{\hbar}\frac{1}{(2\pi\hbar)^{3/2}}\frac{e^{-i\frac{\mathrm{E}}{\hbar}t}}{(2\pi)^{3/4}\sigma_r^{3/2}} p_\rho e^{-\frac{\sigma_r^2}{\hbar^2}p_z^2} \int_0^{\pi/2} e^{-\sigma_r^2 \frac{p_\rho^2}{\hbar^2}\cos^2\phi'} \cos^2\phi' d\phi',$$
$$\int_{\mathbb{R}^3} f_w d^3P = \left|\tilde{\Psi}^{(\mathrm{E})}\right|^2 = \frac{32\sigma_r^5}{\pi(2\pi)^{3/2}\hbar^5} P_\rho^2 e^{-2\frac{\sigma_r^2}{\hbar^2}P_z^2} \left(\int_0^{\pi/2} e^{-\sigma_r^2 \frac{P_\rho^2}{\hbar^2}\cos^2\phi'} \cos^2\phi' d\phi'\right)^2. \qquad \text{(A.26)}$$

Note that momentum representation $\tilde{\Psi}^{(\mathrm{EM})}$ has the form of a Gaussian distribution, that is

$$\left|\tilde{\Psi}^{(\mathrm{EM})}\right|^2 = \left(\frac{\sigma_r\sqrt{2\pi}}{\pi\hbar}\right)^3 e^{-\frac{2\sigma_r^2}{\hbar^2}P^2}, \tag{A.27}$$

and does not coincide with distribution (A.26). For vector potential $\vec{A}_2$, the following expression is valid:

$$\frac{q}{2}\int_{-1}^{1}\vec{s}\cdot\vec{A}_2\left(\vec{r}+\tau\frac{\vec{s}}{2}\right)d\tau = q\vec{s}\cdot\vec{A}_2\left(\vec{r}\right). \tag{A.28}$$

Using expression (A.28), we find function $f_w\left(\vec{r},\vec{P}\right)$ corresponding to wave function $\Psi^{(\mathrm{EM})}$

$$\left(2\pi\hbar\right)^3 f_w\left(\vec{r},\vec{P}\right) = \frac{e^{-\frac{r^2}{2\sigma_r^2}-\frac{2\sigma_r^2}{\hbar^2}\left[\vec{P}+q\vec{A}_2(\vec{r})\right]^2}}{\left(2\pi\right)^{3/2}\sigma_r^3}\int_{\mathbb{R}^3} e^{-\frac{1}{2\sigma_r^2}\left(\frac{\vec{s}}{2}+i2\frac{\sigma_r^2}{\hbar}\left[\vec{P}+q\vec{A}_2(\vec{r})\right]\right)^2} d^3s,$$

$$f_w\left(\vec{r},\vec{P}\right) = \frac{1}{\left(\pi\hbar\right)^3} e^{-\frac{r^2}{2\sigma_r^2}-\frac{2\sigma_r^2}{\hbar^2}\left[\vec{P}+q\vec{A}_2(\vec{r})\right]^2}. \tag{A.29}$$

For the momentum representation, we obtain

$$\int_{\mathbb{R}^3} f_w\left(\vec{r},\vec{P}\right)d^3r = \frac{\sigma_r\sqrt{2\pi}}{\left(\pi\hbar\right)^3} e^{-\frac{2\sigma_r^2}{\hbar^2}P^2}\int_0^{2\pi}d\phi'\int_0^{+\infty} e^{-\left(\frac{1}{2\sigma_r^2}+\frac{2q^2\sigma_r^2c_0^2}{\hbar^2}\right)\rho'^2-\frac{4qc_0\sigma_r^2}{\hbar^2}P_\rho\sin(\phi'-\phi_P)\rho'}\rho' d\rho' =$$
$$= \frac{\sigma_r\sqrt{2\pi}}{\left(\pi\hbar\right)^3 c_1^2} e^{-\frac{2\sigma_r^2}{\hbar^2}P^2}\int_0^{2\pi} e^{\frac{c_2^2}{2\sigma_r^2c_1^2}P_\rho^2\sin^2\phi''}d\phi''\int_0^{+\infty} e^{-\frac{1}{2\sigma_r^2}\left(\rho''+\frac{c_2}{c_1}P_\rho\sin\phi''\right)^2}\rho'' d\rho'', \tag{A.30}$$

where

$$c_0 = -\frac{\hbar\eta}{2q\sigma_r^2},\ c_1^2 = 1+\frac{4q^2\sigma_r^4c_0^2}{\hbar^2},\ c_2 = \frac{4qc_0\sigma_r^4}{\hbar^2},\ \rho'' = c_1\rho'. \tag{A.31}$$

Let us calculate the integral

$$\int_0^{+\infty} e^{-\frac{1}{2\sigma_r^2}(\rho''+\lambda)^2}\rho'' d\rho'' = \sigma_r^2 e^{-\frac{\lambda^2}{2\sigma_r^2}} - \lambda\sigma_r\sqrt{\frac{\pi}{2}}\,\mathrm{erfc}\left(\frac{\lambda}{\sigma_r\sqrt{2}}\right), \tag{A.32}$$

where $\mathrm{erfc}\left(x\right) = \frac{2}{\sqrt{\pi}}\int_x^{+\infty} e^{-t^2}dt$. Substituting (A.32) into expression (A.30), we obtain:

$$\int\limits_{\mathbb{R}^3} f_w\left(\vec{r},\vec{P}\right)d^3r = \frac{\sigma_r\sqrt{2\pi}}{\left(\pi\hbar\right)^3 c_1^2} e^{-\frac{2\sigma_r^2}{\hbar^2}P^2}\int\limits_0^{2\pi} e^{\frac{c_2^2P_\rho^2\sin^2\phi''}{2\sigma_r^2c_1^2}}d\phi''\left(\sigma_r^2 e^{-\frac{c_2^2P_\rho^2\sin^2\phi''}{2c_1^2\sigma_r^2}} - \frac{c_2\sigma_r}{c_1}\sqrt{\frac{\pi}{2}}\operatorname{erfc}\left(\frac{c_2P_\rho\sin\phi''}{c_1\sigma_r\sqrt{2}}\right)P_\rho\sin\phi''\right) =$$

$$= \frac{\sigma_r^3\left(2\pi\right)^{3/2}}{\left(\pi\hbar\right)^3 c_1^2} e^{-\frac{2\sigma_r^2}{\hbar^2}P^2} + \frac{4\sigma_r^2 c_2}{\pi^2\hbar^3 c_1^3} e^{-\frac{2\sigma_r^2}{\hbar^2}P^2} P_\rho \int\limits_0^{\pi/2} e^{\frac{c_2^2P_\rho^2\sin^2\phi''}{2\sigma_r^2c_1^2}} \operatorname{erf}\left(\frac{c_2P_\rho\sin\phi''}{c_1\sigma_r\sqrt{2}}\right)\sin\phi''d\phi'' \qquad \text{(A.33)}$$

with the periodicity property of the integrand is taken into account:

$$\int\limits_0^{2\pi} e^{\frac{c_2^2}{2\sigma_r^2c_1^2}P_\rho^2\sin^2\phi''} \operatorname{erfc}\left(\frac{c_2P_\rho\sin\phi''}{c_1\sigma_r\sqrt{2}}\right)\sin\phi''d\phi'' = -4\int\limits_0^{\pi/2} e^{\frac{c_2^2}{2\sigma_r^2c_1^2}P_\rho^2\sin^2\phi''} \operatorname{erf}\left(\frac{c_2P_\rho\sin\phi''}{c_1\sigma_r\sqrt{2}}\right)\sin\phi''d\phi''.$$

Let us transform the coefficients (A.31) $c_1^2 = 1+\eta^2$, $\frac{c_2}{c_1} = -\frac{2\sigma_r^2\eta}{\hbar\sqrt{1+\eta^2}}$. As a result, expression (A.33) will take the form:

$$\int\limits_{\mathbb{R}^3} f_w d^3r = \frac{2\sigma_r^2 e^{-\frac{2\sigma_r^2}{\hbar^2}P^2}}{\pi^2\hbar^3\left(1+\eta^2\right)}\left[\sigma_r\sqrt{2\pi} - \frac{4\sigma_r^2\eta}{\hbar\sqrt{1+\eta^2}}P_\rho\int\limits_0^{\pi/2} e^{\frac{2\sigma_r^2\eta^2P_\rho^2\sin^2\phi}{\hbar^2\left(1+\eta^2\right)}} \operatorname{erf}\left(-\frac{\sigma_r\eta\sqrt{2}}{\hbar\sqrt{1+\eta^2}}P_\rho\sin\phi\right)\sin\phi d\phi\right].$$

Corollary 3 is proved.

## Appendix B

***Proof of Theorem 5***

Let us find the partial time derivative of the Wigner function of the pure state (i.22):

$$\left(2\pi\hbar\right)^3\frac{\partial}{\partial t}W\left(\vec{r},\vec{p},t\right) = \int\limits_{(\infty)} e^{-i\frac{\vec{s}\cdot\vec{p}}{\hbar}}\Psi\left(\vec{r}'',t\right)\frac{\partial}{\partial t}\bar{\Psi}\left(\vec{r}',t\right)d^3s + \int\limits_{(\infty)} e^{-i\frac{\vec{s}\cdot\vec{p}}{\hbar}}\bar{\Psi}\left(\vec{r}',t\right)\frac{\partial}{\partial t}\Psi\left(\vec{r}'',t\right)d^3s. \qquad \text{(B.1)}$$

It follows from the Schrödinger equation (i.13) that

$$\left(\hat{\mathrm{p}} - q\vec{A}\right)^2\Psi = -\hbar^2\Delta_r\Psi + 2i\hbar q\left(\nabla_r\Psi,\vec{A}\right) + q^2\left|\vec{A}\right|^2\Psi,$$

where the condition $\operatorname{div}_r\vec{A} = 0$ is taken into account. As a result, the Schrödinger equation takes the form:

$$\begin{aligned} i\hbar\frac{\partial}{\partial t}\Psi\left(\vec{r}'',t\right) &= \left[-\frac{\hbar^2}{2m}\Delta_{r''} + i\frac{\hbar q}{m}\vec{A}\left(\vec{r}'',t\right)\cdot\nabla_{r''} + \frac{q^2}{2m}\left|\vec{A}\left(\vec{r}'',t\right)\right|^2 + U\left(\vec{r}'',t\right)\right]\Psi\left(\vec{r}'',t\right), \\ -i\hbar\frac{\partial}{\partial t}\bar{\Psi}\left(\vec{r}',t\right) &= \left[-\frac{\hbar^2}{2m}\Delta_{r'} - i\frac{\hbar q}{m}\vec{A}\left(\vec{r}',t\right)\cdot\nabla_{r'} + \frac{q^2}{2m}\left|\vec{A}\left(\vec{r}',t\right)\right|^2 + U\left(\vec{r}',t\right)\right]\bar{\Psi}\left(\vec{r}',t\right). \end{aligned} \qquad \text{(B.2)}$$

Substituting equations (B.1) into the first and second integrals of expression (B.1), we obtain

$$\left(2\pi\hbar\right)^3 \frac{\partial}{\partial t} W = I_1 + I_2 + I_3 + I_4, \tag{B.3}$$

where

$$I_1 = i\frac{\hbar}{2m}\int_{(\infty)} e^{-i\frac{\vec{s}\cdot\vec{p}}{\hbar}} \left[\overline{\Psi}\left(\vec{r}',t\right)\Delta_{r''}\Psi\left(\vec{r}'',t\right) - \Psi\left(\vec{r}'',t\right)\Delta_{r'}\overline{\Psi}\left(\vec{r}',t\right)\right] d^3 s, \tag{B.4}$$

$$I_2 = \frac{q}{m}\int_{(\infty)} e^{-i\frac{\vec{s}\cdot\vec{p}}{\hbar}} \left[\Psi\left(\vec{r}'',t\right)\vec{A}\left(\vec{r}',t\right)\cdot\nabla_{r'}\overline{\Psi}\left(\vec{r}',t\right) + \overline{\Psi}\left(\vec{r}',t\right)\vec{A}\left(\vec{r}'',t\right)\cdot\nabla_{r''}\Psi\left(\vec{r}'',t\right)\right] d^3 s, \tag{B.5}$$

$$I_3 = i\frac{q^2}{2m\hbar}\int_{(\infty)} e^{-i\frac{\vec{s}\cdot\vec{p}}{\hbar}} \overline{\Psi}\left(\vec{r}',t\right)\Psi\left(\vec{r}'',t\right)\left[\left|\vec{A}\left(\vec{r}',t\right)\right|^2 - \left|\vec{A}\left(\vec{r}'',t\right)\right|^2\right] d^3 s, \tag{B.6}$$

$$I_4 = \frac{i}{\hbar}\int_{(\infty)} e^{-i\frac{\vec{s}\cdot\vec{p}}{\hbar}} \Psi\left(\vec{r}'',t\right)\left[U\left(\vec{r}',t\right) - U\left(\vec{r}'',t\right)\right]\overline{\Psi}\left(\vec{r}',t\right) d^3 s. \tag{B.7}$$

We have calculated integral $I_4$. Let us transform the integrand:

$$U\left(\vec{r}',t\right) - U\left(\vec{r}'',t\right) = -\sum_{l=0}^{+\infty} \frac{\left(\vec{s}\cdot\nabla_r\right)^{2l+1}}{2^{2l}\left(2l+1\right)!} U\left(\vec{r},t\right). \tag{B.8}$$

Let us perform a transormation

$$e^{-i\frac{\vec{s}\cdot\vec{p}}{\hbar}} \left(i\frac{\vec{s}}{\hbar}\cdot\vec{\nabla}_r\right)^{2l+1} U\left(\vec{r},t\right) = -\left(\vec{\nabla}_p\cdot\vec{\nabla}_r\right)^{2l+1} e^{-i\frac{\vec{s}\cdot\vec{p}}{\hbar}} U\left(\vec{r},t\right). \tag{B.9}$$

We substitute expressions (B.8) and (B.9) into the integral for potential energy (B.7), we obtain

$$I_4 = \left(2\pi\hbar\right)^3 \sum_{l=0}^{+\infty} \frac{\left(-1\right)^l \left(\hbar/2\right)^{2l}}{\left(2l+1\right)!} U\left(\vec{r},t\right)\left(\overleftarrow{\nabla}_r\cdot\vec{\nabla}_p\right)^{2l+1} W\left(\vec{r},\vec{p},t\right). \tag{B.10}$$

We transform integral $I_1$ as:

$$I_1 = i\frac{\hbar}{m}\int_{(\infty)} e^{-i\frac{\vec{s}\cdot\vec{p}}{\hbar}} \left(\nabla_r\cdot\nabla_s\right)\overline{\Psi}\left(\vec{r}',t\right)\Psi\left(\vec{r}'',t\right) d^3 s = i\frac{\hbar}{m}\nabla_r\cdot\int_{(\infty)} e^{-i\frac{\vec{s}\cdot\vec{p}}{\hbar}} \nabla_s\overline{\Psi}\left(\vec{r}',t\right)\Psi\left(\vec{r}'',t\right) d^3 s, \tag{B.11}$$

where it is taken into account that $\vec{r} = \frac{\vec{r}' + \vec{r}''}{2}$, $\vec{s} = \vec{r}'' - \vec{r}'$,

$$\nabla_{r'} = \frac{1}{2}\nabla_r - \nabla_s, \qquad \nabla_{r''} = \frac{1}{2}\nabla_r + \nabla_s, \tag{B.12}$$

$$\Delta_{r'} - \Delta_{r''} = -2\left(\nabla_r\cdot\nabla_s\right). \tag{B.13}$$

We transform integranl (B.11)

$$e^{-i\frac{\vec{s}\cdot\vec{p}}{\hbar}}\nabla_s\left[\bar{\Psi}(\vec{r}',t)\Psi(\vec{r}'',t)\right]=\nabla_s\left[e^{-i\frac{\vec{s}\cdot\vec{p}}{\hbar}}\bar{\Psi}(\vec{r}',t)\Psi(\vec{r}'',t)\right]+\frac{i}{\hbar}\vec{p}\bar{\Psi}(\vec{r}',t)\Psi(\vec{r}'',t)e^{-i\frac{\vec{s}\cdot\vec{p}}{\hbar}}. \tag{B.14}$$

Taking into account (B.14), the integral (B.11) takes the form:

$$I_1=-\frac{\vec{p}}{m}\cdot\nabla_r\int_{(\infty)}e^{-i\frac{\vec{s}\cdot\vec{p}}{\hbar}}\bar{\Psi}(\vec{r}',t)\Psi(\vec{r}'',t)d^3s=-(2\pi\hbar)^3\frac{\vec{p}}{m}\cdot\nabla_r W. \tag{B.15}$$

Let us consider integral $I_2$, using relations (B.12):

$$I_2=\frac{q}{2m}\int_{(\infty)}e^{-i\frac{\vec{s}\cdot\vec{p}}{\hbar}}\left[\vec{A}(\vec{r}',t)+\vec{A}(\vec{r}'',t)\right]\cdot\nabla_r\bar{\Psi}(\vec{r}',t)\Psi(\vec{r}'',t)d^3s+$$
$$+\frac{q}{m}\int_{(\infty)}e^{-i\frac{\vec{s}\cdot\vec{p}}{\hbar}}\left[\vec{A}(\vec{r}'',t)-\vec{A}(\vec{r}',t)\right]\cdot\nabla_s\bar{\Psi}(\vec{r}',t)\Psi(\vec{r}'',t)d^3s.$$

Let us take into account relations:

$$e^{-i\frac{\vec{s}\cdot\vec{p}}{\hbar}}\left[\vec{A}(\vec{r}',t)+\vec{A}(\vec{r}'',t)\right]\cdot\nabla_r\bar{\Psi}(\vec{r}',t)\Psi(\vec{r}'',t)=\nabla_r\left\{e^{-i\frac{\vec{s}\cdot\vec{p}}{\hbar}}\left[\vec{A}(\vec{r}',t)+\vec{A}(\vec{r}'',t)\right]\bar{\Psi}(\vec{r}',t)\Psi(\vec{r}'',t)\right\},$$

and

$$e^{-i\frac{\vec{s}\cdot\vec{p}}{\hbar}}\left[\vec{A}(\vec{r}'',t)-\vec{A}(\vec{r}',t)\right]\cdot\nabla_s\bar{\Psi}(\vec{r}',t)\Psi(\vec{r}'',t)=\nabla_s\cdot\left\{e^{-i\frac{\vec{s}\cdot\vec{p}}{\hbar}}\left[\vec{A}(\vec{r}'',t)-\vec{A}(\vec{r}',t)\right]\bar{\Psi}(\vec{r}',t)\Psi(\vec{r}'',t)\right\}+$$
$$+\frac{i}{\hbar}\vec{p}\cdot\left[\vec{A}(\vec{r}'',t)-\vec{A}(\vec{r}',t)\right]\bar{\Psi}(\vec{r}',t)\Psi(\vec{r}'',t)e^{-i\frac{\vec{s}\cdot\vec{p}}{\hbar}}.$$

As a result

$$I_2=\frac{q}{2m}\nabla_r\cdot\int_{(\infty)}e^{-i\frac{\vec{s}\cdot\vec{p}}{\hbar}}\left[\vec{A}(\vec{r}',t)+\vec{A}(\vec{r}'',t)\right]\bar{\Psi}(\vec{r}',t)\Psi(\vec{r}'',t)d^3s+$$
$$+\frac{q}{m}\frac{i}{\hbar}\vec{p}\cdot\int_{(\infty)}\left[\vec{A}(\vec{r}'',t)-\vec{A}(\vec{r}',t)\right]\bar{\Psi}(\vec{r}',t)\Psi(\vec{r}'',t)e^{-i\frac{\vec{s}\cdot\vec{p}}{\hbar}}d^3s=I_{21}+I_{22}. \tag{B.16}$$

Let us use the Taylor's series expansions:

$$\vec{A}(\vec{r}',t)+\vec{A}(\vec{r}'',t)=\sum_{n=0}^{+\infty}\frac{(-1)^n(\vec{s}\cdot\nabla_r)^n}{2^n n!}\vec{A}(\vec{r})+\sum_{n=0}^{+\infty}\frac{(\vec{s}\cdot\nabla_r)^n}{2^n n!}\vec{A}(\vec{r})=$$
$$=\sum_{n=0}^{+\infty}\frac{\left[1+(-1)^n\right](\vec{s}\cdot\nabla_r)^n\vec{A}(\vec{r})}{2^n n!}=2\sum_{k=0}^{+\infty}\frac{(\vec{s}\cdot\nabla_r)^{2k}\vec{A}(\vec{r})}{2^{2k}(2k)!}, \tag{B.17}$$
$$\vec{A}(\vec{r}'',t)-\vec{A}(\vec{r}',t)=\sum_{n=0}^{+\infty}\left[1-(-1)^n\right]\frac{(\vec{s}\cdot\nabla_r)^n}{2^n n!}\vec{A}(\vec{r})=\sum_{k=0}^{+\infty}\frac{(\vec{s}\cdot\nabla_r)^{2k+1}}{2^{2k}(2k+1)!}\vec{A}(\vec{r}).$$

Substituting expansions (B.17) into integral (B.16), we obtain:

$$I_{21} = (2\pi\hbar)^3 \frac{q}{m} \nabla_r \cdot \sum_{k=0}^{+\infty} \frac{(-1)^k (\hbar/2)^{2k}}{(2k)!} \vec{A}(\vec{r},t)\left(\overleftarrow{\nabla}_r \cdot \vec{\nabla}_p\right)^{2k} W(\vec{r},\vec{p},t), \tag{B.18}$$

where it is taken into account that

$$e^{-i\frac{\vec{s}\cdot\vec{p}}{\hbar}} \left(i\frac{\vec{s}}{\hbar}\cdot\nabla_r\right)^{2k} \vec{A}(\vec{r},t) = e^{-i\frac{\vec{s}\cdot\vec{p}}{\hbar}} \left(\overleftarrow{\nabla}_p \cdot \vec{\nabla}_r\right)^{2k} \vec{A}(\vec{r},t). \tag{B.19}$$

Similarly, for integral $I_{22}$ we have

$$I_{22} = -(2\pi\hbar)^3 \frac{q}{m} \vec{p} \cdot \sum_{k=0}^{+\infty} \frac{(-1)^k (\hbar/2)^{2k}}{(2k+1)!} W(\vec{r},\vec{p},t)\left(\overleftarrow{\nabla}_p \cdot \vec{\nabla}_r\right)^{2k+1} \vec{A}(\vec{r},t), \tag{B.20}$$

which takes into account that

$$e^{-i\frac{\vec{s}\cdot\vec{p}}{\hbar}} \left(i\frac{\vec{s}}{\hbar}\cdot\nabla_r\right)^{2k+1} \vec{A}(\vec{r}) = -e^{-i\frac{\vec{s}\cdot\vec{p}}{\hbar}} \left(\overleftarrow{\nabla}_p \cdot \vec{\nabla}_r\right)^{2k+1} \vec{A}(\vec{r},t). \tag{B.21}$$

Substituting (B.20) and (B.18) into expression (B.16), we obtain

$$\frac{m}{q}\frac{I_2}{(2\pi\hbar)^3} = \nabla_r \cdot \sum_{k=0}^{+\infty} \frac{(-1)^k (\hbar/2)^{2k}}{(2k)!} \vec{A}\left(\overleftarrow{\nabla}_r \cdot \vec{\nabla}_p\right)^{2k} W - \vec{p} \cdot \sum_{k=0}^{+\infty} \frac{(-1)^k (\hbar/2)^{2k}}{(2k+1)!} \vec{A}\left(\overleftarrow{\nabla}_r \cdot \vec{\nabla}_p\right)^{2k+1} W. \tag{B.22}$$

Let us find integral $I_3$ using expansions (B.17):

$$\frac{m}{q^2}\frac{1}{(2\pi\hbar)^3} I_3 = \tag{B.23}$$

$$= \sum_{l=0}^{+\infty} \frac{(-1)^l (\hbar/2)^{2l}}{(2l+1)!} \sum_{k=0}^{+\infty} \frac{(-1)^k (\hbar/2)^{2k}}{(2k)!} \vec{A}(\vec{r},t)\left(\overleftarrow{\nabla}_r \cdot \vec{\nabla}_p\right)^{2k} \left[W(\vec{r},\vec{p},t)\right]\left(\overleftarrow{\nabla}_p \cdot \vec{\nabla}_r\right)^{2l+1} \vec{A}(\vec{r},t).$$

As a result

$$\frac{\partial}{\partial t}W + \frac{\vec{p}}{m}\cdot\nabla_r W = \sum_{l=0}^{+\infty} \frac{(-1)^l (\hbar/2)^{2l}}{(2l+1)!} U\left(\overleftarrow{\nabla}_r \cdot \vec{\nabla}_p\right)^{2l+1} W + \frac{q}{m}\nabla_r \sum_{k=0}^{+\infty} \frac{(-1)^k (\hbar/2)^{2k}}{(2k)!} \vec{A}\left(\overleftarrow{\nabla}_r \cdot \vec{\nabla}_p\right)^{2k} W$$

$$-\frac{q}{m}\vec{p}\sum_{k=0}^{+\infty} \frac{(-1)^k (\hbar/2)^{2k}}{(2k+1)!} \vec{A}\left(\overleftarrow{\nabla}_r \cdot \vec{\nabla}_p\right)^{2k+1} W + \tag{B.24}$$

$$+\frac{q^2}{m}\sum_{l=0}^{+\infty} \frac{(-1)^l (\hbar/2)^{2l}}{(2l+1)!} \sum_{k=0}^{+\infty} \frac{(-1)^k (\hbar/2)^{2k}}{(2k)!} \vec{A}(\vec{r},t)\left(\overleftarrow{\nabla}_r \cdot \vec{\nabla}_p\right)^{2k} \left[W(\vec{r},\vec{p},t)\right]\left(\overleftarrow{\nabla}_p \cdot \vec{\nabla}_r\right)^{2l+1} \vec{A}(\vec{r},t)$$

We transform obtained equation (B.24)

$$\frac{\partial}{\partial t}W+\frac{1}{m}\left\{\vec p-q\sum_{k=0}^{+\infty}\frac{(-1)^k(\hbar/2)^{2k}}{(2k)!}\vec A\left(\overleftarrow{\nabla}_r\cdot\overrightarrow{\nabla}_p\right)^{2k}\right\}\vec\nabla_r W=\left(\vec\nabla_p W\cdot\vec\nabla_r\right)\sum_{l=0}^{+\infty}\frac{(-1)^l(\hbar/2)^{2l}}{(2l+1)!}\left(\overleftarrow{\nabla}_p\cdot\overrightarrow{\nabla}_r\right)^{2l}U-$$

$$-\frac{q}{m}\left\{\vec p-q\sum_{k=0}^{+\infty}\frac{(-1)^k(\hbar/2)^{2k}}{(2k)!}\vec A\left(\overleftarrow{\nabla}_r\cdot\overrightarrow{\nabla}_p\right)^{2k}\right\}\left(\vec\nabla_p W\cdot\vec\nabla_r\right)\sum_{l=0}^{+\infty}\frac{(-1)^l(\hbar/2)^{2l}}{(2l+1)!}\left(\overleftarrow{\nabla}_p\cdot\overrightarrow{\nabla}_r\right)^{2l}\vec A=0, \quad \text{(B.25)}$$

where operator $\nabla_p$ acts only on the Wigner function. Let us transform the last term in equation (B.25)

$$\left[p_\beta-q\sum_{l=0}^{+\infty}\frac{(-1)^l(\hbar/2)^{2l}}{(2l+1)!}A_\beta\left(\overleftarrow{\nabla}_r\cdot\overrightarrow{\nabla}_p\right)^{2l}\right]\frac{\partial W}{\partial p_\alpha}\sum_{k=0}^{+\infty}\frac{(-1)^k(\hbar/2)^{2k}}{(2k+1)!}\left(\overleftarrow{\nabla}_p\cdot\overrightarrow{\nabla}_r\right)^{2k}\frac{\partial A_\beta}{\partial x_\alpha}=$$

$$=\sum_{k=0}^{+\infty}\frac{(-1)^k(\hbar/2)^{2k}}{(2k+1)!}\frac{\partial A_\beta}{\partial x_\alpha}\left(\overleftarrow{\nabla}_r\cdot\overrightarrow{\nabla}_p\right)^{2k}\frac{\partial W}{\partial p_\alpha}\left[p_\beta-q\sum_{l=0}^{+\infty}\frac{(-1)^l(\hbar/2)^{2l}}{(2l+1)!}\left(\overleftarrow{\nabla}_p\cdot\overrightarrow{\nabla}_r\right)^{2l}A_\beta\right]. \quad \text{(B.26)}$$

We break expression (B.26) into two identical terms:

$$-q\left[p_\beta-q\sum_{l=0}^{+\infty}\frac{(-1)^l(\hbar/2)^{2l}}{(2l+1)!}A_\beta\left(\overleftarrow{\nabla}_r\cdot\overrightarrow{\nabla}_p\right)^{2l}\right]\frac{\partial W}{\partial p_\alpha}\sum_{k=0}^{+\infty}\frac{(-1)^k(\hbar/2)^{2k}}{(2k+1)!}\left(\overleftarrow{\nabla}_p\cdot\overrightarrow{\nabla}_r\right)^{2k}\frac{\partial A_\beta}{\partial x_\alpha}=$$

$$=\frac{1}{2}\left(\nabla_p W\cdot\vec\nabla_r\right)\left[\vec p-q\sum_{l=0}^{+\infty}\frac{(-1)^l(\hbar/2)^{2l}}{(2l+1)!}\left(\overleftarrow{\nabla}_p\cdot\overrightarrow{\nabla}_r\right)^{2l}\vec A\right]^2. \quad \text{(B.27)}$$

Substituting expression (B.27) into equation (B.25), we obtain

$$\frac{\partial}{\partial t}W+\frac{1}{m}\left\{\vec p-q\sum_{k=0}^{+\infty}\frac{(-1)^k(\hbar/2)^{2k}}{(2k)!}\vec A\left(\overleftarrow{\nabla}_r\cdot\overrightarrow{\nabla}_p\right)^{2k}\right\}\nabla_r W-$$
$$=\left(\vec\nabla_p W\cdot\nabla_r\right)\left\{\frac{1}{2m}\left[\vec p-q\sum_{l=0}^{+\infty}\frac{(-1)^l(\hbar/2)^{2l}}{(2l+1)!}\left(\overleftarrow{\nabla}_p\cdot\overrightarrow{\nabla}_r\right)^{2l}\vec A\right]^2+\sum_{l=0}^{+\infty}\frac{(-1)^l(\hbar/2)^{2l}}{(2l+1)!}\left(\overleftarrow{\nabla}_p\cdot\overrightarrow{\nabla}_r\right)^{2l}U\right\}=0. \quad \text{(B.28)}$$

Using the notation (2.4)-(2.5) of operators $q\hat{\mathcal{U}}$ and $\hat{\mathcal{A}}$, equation (B.28) takes the form

$$\frac{\partial}{\partial t}W+\frac{1}{m}\left(\vec p-q\hat{\vec{\mathcal{A}}}\right)\cdot\nabla_r W-\left(\nabla_p W\cdot\nabla_r\right)\left[\frac{1}{2m}\left(\vec p-q\hat{\overleftarrow{\mathcal{A}}}\right)^2+\hat{\overleftarrow{\mathcal{U}}}\right]=0. \quad \text{(B.29)}$$

Then we reduce equation (B.29) to the second Vlasov equation.

$$\frac{\partial}{\partial t}W+\frac{1}{m}\mathrm{div}_r\left[W\left(\vec p-q\hat{\overleftarrow{\mathcal{A}}}\right)\right]-\left\{W\,\mathrm{div}_r\frac{1}{m}\left(\vec p-q\hat{\overleftarrow{\mathcal{A}}}\right)+\left(\nabla_p W\cdot\nabla_r\right)\left[\frac{1}{2m}\left(\vec p-q\hat{\overleftarrow{\mathcal{A}}}\right)^2+\hat{\overleftarrow{\mathcal{U}}}\right]\right\}=0. \quad \text{(B.30)}$$

Let us transform the last term in equation (B.30) and perform intermediate calculations:

$$\operatorname{div}_p\left\{W\nabla_r\left[\frac{1}{2m}\left(\vec{p}-q\overleftarrow{\hat{\mathcal{A}}}\right)^2+\overleftarrow{\hat{\mathcal{U}}}\right]\right\}=\nabla_p W\cdot\nabla_r\left[\frac{1}{2m}\left(\vec{p}-q\overleftarrow{\hat{\mathcal{A}}}\right)^2+\overleftarrow{\hat{\mathcal{U}}}\right]+ \\ +W\operatorname{div}_p\left\{\nabla_r\left[\frac{1}{2m}\left(\vec{p}-q\overleftarrow{\hat{\mathcal{A}}}\right)^2+\overleftarrow{\hat{\mathcal{U}}}\right]\right\}. \tag{B.31}$$

We keep in mind that

$$\operatorname{div}_p\left\{\nabla_r\left[\frac{1}{2m}\left(\vec{p}-q\overleftarrow{\hat{\mathcal{A}}}\right)^2+\overleftarrow{\hat{\mathcal{U}}}\right]\right\}=\frac{1}{m}\operatorname{div}_r\left(\vec{p}-q\overleftarrow{\hat{\mathcal{A}}}\right). \tag{B.32}$$

Substituting (B.32) into (B.31), we derive

$$\operatorname{div}_p\left\{W\nabla_r\left[\frac{1}{2m}\left(\vec{p}-q\overleftarrow{\hat{\mathcal{A}}}\right)^2+\overleftarrow{\hat{\mathcal{U}}}\right]\right\}=\nabla_p W\cdot\nabla_r\left[\frac{1}{2m}\left(\vec{p}-q\overleftarrow{\hat{\mathcal{A}}}\right)^2+\overleftarrow{\hat{\mathcal{U}}}\right]+W\frac{1}{m}\operatorname{div}_r\left(\vec{p}-q\overleftarrow{\hat{\mathcal{A}}}\right). \tag{B.33}$$

Taking (B.33) into account, equation (B.30) takes the form

$$\frac{\partial}{\partial t}W+\frac{1}{m}\operatorname{div}_r\left[\left(\vec{p}-q\vec{\hat{\mathcal{A}}}\right)W\right]+\operatorname{div}_p\left\{-\nabla_r\left[\frac{1}{2m}\left(\vec{p}-q\vec{\hat{\mathcal{A}}}\right)^2+\vec{\hat{\mathcal{U}}}\right]W\right\}=0. \tag{B.34}$$

Theorem 5 is proved.

***Proof of Theorem 6***

Let us write equation (2.1) with respect to the Vlasov function $f_V\left(\vec{r},\vec{P},t\right)$. The derivative of operator $\overleftarrow{\hat{\mathcal{H}}}$ takes the form:

$$\frac{\partial}{\partial x_\beta}\overleftarrow{\hat{\mathcal{H}}}=\frac{\partial}{\partial x_\beta}\left(\frac{1}{2m}\overleftarrow{\hat{\mathcal{P}}}^2+\overleftarrow{\hat{\mathcal{U}}}\right)=\frac{1}{m}\overleftarrow{\hat{\mathcal{P}}}_\mu\frac{\partial\overleftarrow{\hat{\mathcal{P}}}_\mu}{\partial x_\beta}+\frac{\partial\overleftarrow{\hat{\mathcal{U}}}}{\partial x_\beta}=-q\frac{1}{m}\overleftarrow{\hat{\mathcal{P}}}_\mu\frac{\partial\overleftarrow{\hat{\mathcal{A}}}_\mu}{\partial x_\beta}+\frac{\partial\overleftarrow{\hat{\mathcal{U}}}}{\partial x_\beta}. \tag{B.35}$$

Note that, according to the definition (2.4)-(2.5), operator $\overleftarrow{\hat{\mathcal{A}}}_\mu$ contains the derivative with respect to variable $\vec{p}$. When passing from variable $\vec{p}$ to variable $\vec{P}$, the form of operator $\overleftarrow{\hat{\mathcal{A}}}_\mu$ does not change, since $\vec{P}=\vec{p}-q\vec{A}$ and $\overleftarrow{\nabla}_p=\overleftarrow{\nabla}_P$. Equation (B.29) is written with respect to the Wigner function $W\left(\vec{r},\vec{p},t\right)$, so let us rewrite it for function $W\left(\vec{r},\vec{p},t\right)=W\left(\vec{r},\vec{P}+q\vec{A},t\right)=f_V\left(\vec{r},\vec{P},t\right)$, where $\vec{P}=\vec{p}-q\vec{A}\left(\vec{r},t\right)$. We obtain expressions for partial derivatives:

$$\frac{\partial}{\partial t}W=\frac{\partial f_V}{\partial t}-q\frac{\partial f_V}{\partial P_\beta}\frac{\partial A_\beta}{\partial t}, \tag{B.36}$$

$$\frac{\partial}{\partial p_\beta}W=\frac{\partial f_V}{\partial P_\beta}, \tag{B.37}$$

$$\frac{\partial}{\partial x_\beta} W = \frac{\partial f_V}{\partial x_\beta} - q \frac{\partial f_V}{\partial P_\alpha} \frac{\partial A_\alpha}{\partial x_\beta}, \tag{B.38}$$

Substituting expressions (B.36)-(B.38) and (B.35) into equation (2.1), we obtain

$$\frac{\partial f_V}{\partial t} + \frac{1}{m} \vec{\hat{\mathcal{P}}}_\beta \frac{\partial f_V}{\partial x_\beta} - \frac{\partial f_V}{\partial P_\beta} \left( \frac{q}{m} \frac{\partial A_\beta}{\partial x_\alpha} \overleftarrow{\hat{\mathcal{P}}}_\alpha - \frac{q}{m} \overleftarrow{\hat{\mathcal{P}}}_\mu \frac{\partial \overleftarrow{\hat{\mathcal{A}}}_\mu}{\partial x_\beta} + q \frac{\partial A_\beta}{\partial t} + \frac{\partial \bar{\hat{\mathcal{U}}}}{\partial x_\beta} \right) = 0. \tag{B.39}$$

Let us transform the expression in brackets in equation (B.39). According to (2.5), vector potential $\vec{A}$ can be represented as

$$\vec{A} = \overleftarrow{\hat{\mathcal{A}}} - \overleftarrow{\hat{\mathcal{A}}}^{(\hbar)}, \quad \overleftarrow{\hat{\mathcal{A}}}^{(\hbar)} \stackrel{\text{det}}{=} \sum_{k=1}^{+\infty} \frac{(-1)^k (\hbar/2)^{2k}}{(2k)!} \left( \overleftarrow{\nabla}_P \cdot \vec{\nabla}_r \right)^{2k} \vec{A}, \tag{B.40}$$

as a result

$$\frac{\partial A_\beta}{\partial x_\alpha} \overleftarrow{\hat{\mathcal{P}}}_\alpha - \overleftarrow{\hat{\mathcal{P}}}_\mu \frac{\partial \overleftarrow{\hat{\mathcal{A}}}_\mu}{\partial x_\beta} = -\left( \overleftarrow{\hat{\mathcal{P}}} \times \overleftarrow{\hat{\mathcal{B}}} \right)_\beta - \overleftarrow{\hat{\mathcal{P}}}_\alpha \frac{\partial \overleftarrow{\hat{\mathcal{A}}}_\beta^{(\hbar)}}{\partial x_\alpha}, \tag{B.41}$$

which takes into account that $\frac{\partial A_\beta}{\partial x_\alpha} \overleftarrow{\hat{\mathcal{P}}}_\alpha = \overleftarrow{\hat{\mathcal{P}}}_\alpha \frac{\partial A_\beta}{\partial x_\alpha}$, since operator $\overleftarrow{\hat{\mathcal{P}}}_\alpha$ is expressed in terms of operator $\overleftarrow{\hat{\mathcal{A}}}_\alpha$, in which the differentiation with respect to variable $P_\mu$ is performed, and potential $A_\beta$ does not depend on $P_\mu$. Substituting expression (B.41) into equation (B.39), we obtain

$$\frac{\partial f_V}{\partial t} + \frac{1}{m} \vec{\hat{\mathcal{P}}}_\beta \frac{\partial f_V}{\partial x_\beta} + q \frac{\partial f_V}{\partial P_\beta} \left( \frac{1}{m} \left( \overleftarrow{\hat{\mathcal{P}}} \times \overleftarrow{\hat{\mathcal{B}}} \right)_\beta + \frac{\partial \overleftarrow{\hat{\mathcal{A}}}_\beta^{(\hbar)}}{\partial t} + \overleftarrow{\hat{\mathcal{V}}}_\alpha \frac{\partial \overleftarrow{\hat{\mathcal{A}}}_\beta^{(\hbar)}}{\partial x_\alpha} - \frac{\partial \overleftarrow{\hat{\mathcal{A}}}_\beta}{\partial t} - \frac{1}{q} \frac{\partial \bar{\hat{\mathcal{U}}}}{\partial x_\beta} \right) = 0. \tag{B.42}$$

Let us transform the terms in equation (B.42)

$$\vec{\hat{\mathcal{P}}}_\beta \frac{\partial f_V}{\partial x_\beta} = \frac{\partial}{\partial x_\beta} \left[ \left( P_\beta - q \vec{\hat{\mathcal{A}}}_\beta^{(\hbar)} \right) f_V \right], \tag{B.43}$$

where it is taken into account that

$$\frac{\partial}{\partial x_\beta} \left[ A_\beta \left( \overleftarrow{\nabla}_r \cdot \vec{\nabla}_p \right)^{2k} f_V \right] = A_\beta \left( \overleftarrow{\nabla}_r \cdot \vec{\nabla}_p \right)^{2k} \frac{\partial f_V}{\partial x_\beta},$$

since $\operatorname{div}_r \vec{A} = 0$. Similarly for the remaining terms, we obtain:

$$\frac{\partial f_V}{\partial P_\beta} \frac{\partial \bar{\hat{\mathcal{U}}}}{\partial x_\beta} = \frac{\partial}{\partial P_\beta} \left( f_V \frac{\partial \bar{\hat{\mathcal{U}}}}{\partial x_\beta} \right), \tag{B.44}$$

$$\frac{\partial}{\partial P_\beta}\left( f_V \frac{\partial \overleftarrow{\hat{\mathcal{A}}}_\beta}{\partial t} \right) = \frac{\partial f_V}{\partial P_\beta}\frac{\partial \overleftarrow{\hat{\mathcal{A}}}_\beta}{\partial t}, \tag{B.45}$$

$$\frac{\partial}{\partial P_\beta}\left( f_V \frac{\partial \overleftarrow{\hat{\mathcal{A}}}_\beta^{(\hbar)}}{\partial t} \right) = \frac{\partial f_V}{\partial P_\beta}\frac{\partial \overleftarrow{\hat{\mathcal{A}}}_\beta^{(\hbar)}}{\partial t}, \tag{B.46}$$

$$\frac{\partial f_V}{\partial P_\beta}\left( \overleftarrow{\hat{\mathcal{V}}}_\alpha \frac{\partial \overleftarrow{\hat{\mathcal{A}}}_\beta^{(\hbar)}}{\partial x_\alpha} \right) = \frac{\partial}{\partial P_\beta}\left[ f_V \overleftarrow{\hat{\mathcal{V}}}_\alpha \frac{\partial \overleftarrow{\hat{\mathcal{A}}}_\beta^{(\hbar)}}{\partial x_\alpha} \right], \tag{B.47}$$

where $\operatorname{div}_r \vec{A} = 0$ and $P_\alpha \frac{\partial f_V}{\partial P_\beta} = \frac{\partial}{\partial P_\beta}\left( P_\alpha f_V \right) - \delta_{\alpha\beta} f_V$. The last term will take the form:

$$\frac{\partial f_V}{\partial P_\beta}\left( \overleftarrow{\hat{\mathcal{P}}} \times \overleftarrow{\hat{\mathcal{B}}} \right)_\beta = \frac{\partial}{\partial P_\beta}\left[ f_V \left( \overleftarrow{\hat{\mathcal{P}}} \times \overleftarrow{\hat{\mathcal{B}}} \right)_\beta \right], \tag{B.48}$$

where it is taken into account that

$$\frac{\partial f_V}{\partial P_\beta}\left( P_\alpha \overleftarrow{\hat{\mathcal{B}}}_\lambda \right) = \frac{\partial}{\partial P_\beta}\left( f_V P_\alpha \overleftarrow{\hat{\mathcal{B}}}_\lambda \right) - f_V \delta_{\alpha\beta} \overleftarrow{\hat{\mathcal{B}}}_\lambda .$$

Substituting expressions (B.43)-(B.48) into equation (B.42), we derive

$$\frac{\partial f_V}{\partial t} + \frac{1}{m}\frac{\partial}{\partial x_\beta}\left( \overleftarrow{\hat{\mathcal{P}}}_\beta f_V \right) + q\frac{\partial}{\partial P_\beta}\left\{ f_V \left[ \frac{1}{m}\left( \overleftarrow{\hat{\mathcal{P}}} \times \overleftarrow{\hat{\mathcal{B}}} \right)_\beta + \frac{\partial \overleftarrow{\hat{\mathcal{A}}}_\beta^{(\hbar)}}{\partial t} + \overleftarrow{\hat{\mathcal{V}}}_\alpha \frac{\partial \overleftarrow{\hat{\mathcal{A}}}_\beta^{(\hbar)}}{\partial x_\alpha} - \frac{\partial \overleftarrow{\hat{\mathcal{A}}}_\beta}{\partial t} - \frac{1}{q}\frac{\partial \overleftarrow{\hat{\mathcal{U}}}}{\partial x_\beta} \right] \right\} = 0. \tag{B.49}$$

Theorem 6 is proved.

***Proof of Theorem 7***

Integrating the Vlasov-Moyal approximation (2.10) over momentum space, we get

$$f^1 \left\langle \left\langle \dot{\vec{P}} \right\rangle \right\rangle = q \int_{\mathbb{R}^3} f_V \overleftarrow{\hat{\mathcal{E}}} d^3P + q \int_{\mathbb{R}^3} f_V \overleftarrow{\hat{\mathcal{V}}} \times \overleftarrow{\hat{\mathcal{B}}} d^3P + q \int_{\mathbb{R}^3} f_V \frac{\overleftarrow{\mathrm{d}}}{\mathrm{dt}} \overleftarrow{\hat{\mathcal{A}}}^{(\hbar)} d^3P. \tag{B.50}$$

Let us calculate each integral separately:

$$\int_{\mathbb{R}^3} f_V \overleftarrow{\hat{\mathcal{E}}}_\beta d^3P = f^1 \left( -\frac{\partial A_\beta}{\partial t} - \frac{1}{q}\frac{\partial U}{\partial x_\beta} \right). \tag{B.51}$$

$$\int_{\mathbb{R}^3} f_V \left( \overleftarrow{\hat{\mathcal{V}}} \times \overleftarrow{\hat{\mathcal{B}}} \right)_\beta d^3P = \frac{1}{m} f^1 \left\langle \vec{P} \right\rangle_1 \times \vec{B}, \tag{B.52}$$

where $\varepsilon_{\lambda\beta\alpha}\varepsilon_{\lambda\mu\eta} = \delta_{\beta\mu}\delta_{\alpha\eta} - \delta_{\beta\eta}\delta_{\alpha\mu}$ and

$$\int_{\mathbb{R}^3} f_V P_\alpha \left(\overleftarrow{\nabla}_P \cdot \vec{\nabla}_r\right)^{2k} G_{\alpha\beta} d^3P = 0. \tag{B.53}$$

The validity of expression (B.53) results from the following calculations:

$$\begin{aligned}\int_{\mathbb{R}^3} f_V P_\alpha \left(\overleftarrow{\nabla}_P \cdot \vec{\nabla}_r\right)^2 G_{\alpha\beta} d^3P = \int_{\mathbb{R}^3} f_V P_\alpha \left(\overleftarrow{\partial}^2_{P_x}\vec{\partial}^2_x + \overleftarrow{\partial}^2_{P_y}\vec{\partial}^2_y + \overleftarrow{\partial}^2_{P_z}\vec{\partial}^2_z\right) G_{\alpha\beta} d^3P + \\ + 2\int_{\mathbb{R}^3} f_V P_\alpha \left(\overleftarrow{\partial}_{P_x}\overleftarrow{\partial}_{P_y}\vec{\partial}_x\vec{\partial}_y + \overleftarrow{\partial}_{P_x}\overleftarrow{\partial}_{P_z}\vec{\partial}_x\vec{\partial}_z + \overleftarrow{\partial}_{P_y}\overleftarrow{\partial}_{P_z}\vec{\partial}_y\vec{\partial}_z\right) G_{\alpha\beta} d^3P.\end{aligned} \tag{B.54}$$

For terms $\overleftarrow{\partial}^2_{P_\lambda}\vec{\partial}^2_\lambda$ from (B.54), we can get

$$\begin{aligned}&\left(f_V P_\alpha\right)\overleftarrow{\partial}^2_{P_\lambda}\vec{\partial}^2_\lambda G_{\alpha\beta} = \left(P_\alpha \partial^2_{P_\lambda} f_V + 2\delta_{\alpha\lambda}\partial_{P_\lambda} f_V\right)\partial^2_\lambda G_{\alpha\beta}, \\ &\int_{\mathbb{R}^3} \left(f_V P_\alpha\right)\overleftarrow{\partial}^2_{P_\lambda}\vec{\partial}^2_\lambda G_{\alpha\beta} d^3P = \left( P_\alpha \partial_{P_\lambda} f_V\Big|_{\pm\infty} - \delta_{\alpha\lambda}\int_{\mathbb{R}^3} \partial_{P_\lambda} f_V d^3P \right)\partial^2_\lambda G_{\alpha\beta} = 0.\end{aligned} \tag{B.55}$$

And for terms $\overleftarrow{\partial}_{P_\lambda}\overleftarrow{\partial}_{P_\mu}\vec{\partial}_\lambda\vec{\partial}_\mu$ from (B.54), we obtain

$$\begin{aligned}&\left(f_V P_\alpha\right)\overleftarrow{\partial}_{P_\lambda}\overleftarrow{\partial}_{P_\mu}\vec{\partial}_\lambda\vec{\partial}_\mu G_{\alpha\beta} = \left(P_\alpha \partial_{P_\lambda}\partial_{P_\mu} f_V + \delta_{\alpha\lambda}\partial_{P_\mu} f_V + \delta_{\alpha\mu}\partial_{P_\lambda} f_V\right)\vec{\partial}_\lambda\vec{\partial}_\mu G_{\alpha\beta}, \\ &\int_{\mathbb{R}^3} \left(f_V P_\alpha\right)\overleftarrow{\partial}_{P_\lambda}\overleftarrow{\partial}_{P_\mu}\vec{\partial}_\lambda\vec{\partial}_\mu G_{\alpha\beta} d^3P = \left( P_\alpha \partial_{P_\mu} f_V\Big|_{\pm\infty} - \delta_{\alpha\lambda}\int_{\mathbb{R}^3} \partial_{P_\mu} f_V d^3P \right)\vec{\partial}_\lambda\vec{\partial}_\mu G_{\alpha\beta} = 0.\end{aligned} \tag{B.56}$$

The substitution of (B.55) and (B.56) into the integral (B.54) results in zero. A similar result is also valid for higher degrees $\left(\overleftarrow{\nabla}_P \cdot \vec{\nabla}_r\right)^{2k}$. Therefore, expression (B.53) is true. For the last term from expression (B.50), we obtain:

$$\int_{\mathbb{R}^3} f_V \frac{\overleftarrow{\mathrm{d}}}{\mathrm{dt}} \overleftarrow{\bar{\mathcal{A}}}^{(\hbar)}_\beta d^3P = \frac{1}{m}\int_{\mathbb{R}^3} f_V P_\alpha \frac{\partial \overleftarrow{\bar{\mathcal{A}}}^{(\hbar)}_\beta}{\partial x_\alpha} d^3P - \frac{q}{m}\int_{\mathbb{R}^3} f_V \overleftarrow{\bar{\mathcal{A}}}^{(\hbar)}_\alpha \frac{\partial \overleftarrow{\bar{\mathcal{A}}}^{(\hbar)}_\beta}{\partial x_\alpha} d^3P = 0, \tag{B.57}$$

where previous calculations are taken into consideration. Substituting expressions (B.51), (B.52) and (B.57) into the original integral (B.50), we get

$$f^1 \left\langle\left\langle \dot{\vec{P}} \right\rangle\right\rangle = q f^1 \left( -\frac{\partial \vec{A}}{\partial t} - \frac{1}{q}\nabla_r U \right) + \frac{q}{m} f^1 \left\langle \vec{P} \right\rangle \times \vec{B} \tag{B.58}$$

with

$$\int_{\mathbb{R}^3} \vec{\mathcal{P}} f_V d^3P = \int_{\mathbb{R}^3} \vec{P} f_V d^3P - q\int_{\mathbb{R}^3} \vec{\bar{\mathcal{A}}}^{(\hbar)} f_V d^3P = f^1 \left\langle \vec{P} \right\rangle.$$

Theorem 7 is proved.

**Appendix C**

Let us find an expression for the Wigner function $W^{(\mathrm{EM})}$

$$\left(2\pi\hbar\right)^3\left(2\pi\right)^{3/2}\sigma_r^3 W^{(\mathrm{EM})}=e^{-\frac{r^2}{2\sigma_r^2}}\int\limits_{\mathbb{R}^3} e^{-\frac{s^2}{8\sigma_r^2}-\frac{i}{\hbar}\vec{s}\cdot\vec{p}}d^3s=8\left(2\pi\right)^{3/2}\sigma_r^3 e^{-\frac{r^2}{2\sigma_r^2}-\frac{2\sigma_r^2}{\hbar^2}p^2}. \tag{C.1}$$

Expression (C.1) implies the validity of the representation of the Wigner function (1.21).

Let us calculate the value of mean energy $\langle\langle\mathcal{E}\rangle\rangle$ for functions $W^{(\mathrm{EM})}$ and $f_w$. Let us start with function $W^{(\mathrm{EM})}$.

$$\langle\langle\mathcal{E}\rangle\rangle^{(\mathrm{EM})}=\frac{1}{2m\left(\pi\hbar\right)^3}\int\limits_{\mathbb{R}^3} e^{-\frac{r^2}{2\sigma_r^2}}d^3r\int\limits_{\mathbb{R}^3}\left(p^2-2q\vec{p}\cdot\vec{A}+q^2\left|\vec{A}\right|^2\right)e^{-\frac{2\sigma_r^2}{\hbar^2}p^2}d^3p+$$
$$+\frac{\left(2\pi\right)^{3/2}}{\left(\pi\hbar\right)^3}\frac{\hbar^3}{8\sigma_r^3}\frac{\hbar^2}{8m\sigma_r^4}\int\limits_{\mathbb{R}^3}\left(r^2-\frac{4\sigma_r^4}{r^2\sin^2\theta}\right)e^{-\frac{r^2}{2\sigma_r^2}}d^3r=I_T^{(\mathrm{EM})}+I_U^{(\mathrm{EM})}. \tag{C.2}$$

Let us find integrals $I_T^{(\mathrm{EM})}$, $I_U^{(\mathrm{EM})}$ separately. For kinetic integral $I_T^{(\mathrm{EM})}$, we obtain:

$$I_T^{(\mathrm{EM})}=\frac{3\hbar^2}{8m\sigma_r^2}+\frac{\hbar^2}{2m\sigma_r^2}\int\limits_0^{\pi}\frac{d\theta}{\sin\theta}. \tag{C.3}$$

Integral $I_U^{(\mathrm{EM})}$ corresponding to the potential energy has the form:

$$I_U^{(\mathrm{EM})}=\frac{3\hbar^2}{8m\sigma_r^2}-\frac{\hbar^2}{2m\sigma_r^2}\int\limits_0^{\pi}\frac{d\theta}{\sin\theta}. \tag{C.4}$$

Substituting expressions (C.3) and (C.4) into (C.2), we obtain:

$$\langle\langle\mathcal{E}\rangle\rangle^{(\mathrm{EM})}=\int\limits_{\mathbb{R}^3}\int\limits_{\mathbb{R}^3}\left(\frac{P^2}{2m}+U\right)W^{(\mathrm{EM})}d^3r d^3p=\frac{3\hbar^2}{4m\sigma_r^2}. \tag{C.5}$$

Let us perform averaging over function $f_w$.

$$\begin{aligned}\langle\langle\mathcal{E}\rangle\rangle^{(f_w)}&=\int\limits_{\mathbb{R}^3}\int\limits_{\mathbb{R}^3}\left(\frac{P^2}{2m}+U\right)f_w d^3r d^3P=\\&=\frac{1}{2m}\int\limits_{\mathbb{R}^3}d^3r\int\limits_{\mathbb{R}^3}P^2 f_w d^3P+\int\limits_{\mathbb{R}^3}Ud^3r\int\limits_{\mathbb{R}^3}f_w d^3P=I_T^{(f)}+I_U^{(f)}.\end{aligned} \tag{C.6}$$

In order to calculate integral $I_T^{(f)}$, we use the result of [25] where we found the diagonal elements of the pressure tensor for distribution $W^{(\mathrm{E})}$ (i.11). According to Theorem 4, the transition $f_w\left(\vec{r},\vec{P}\right)=W^{(\mathrm{E})}\left(\rho,z,\vec{P}\right)$ is valid, therefore

$$\int\limits_{(\infty)}\left(P_\alpha-\langle P_\alpha\rangle\right)^2 f_w d^3P=f\left(\vec{r}\right)\frac{\hbar^2}{4\sigma_r^2}, \tag{C.7}$$

from here

$$\int_{(\infty)} P_\alpha^2 f_w d^3P = \int_{(\infty)} \left(P_\alpha - \langle P_\alpha \rangle\right)^2 f_w d^3P + f(\vec{r})\left|\langle P_\alpha \rangle\right|^2 = f(\vec{r})\left[\frac{\hbar^2}{4\sigma_r^2} + q^2 A_\alpha^2\right]. \tag{C.8}$$

Taking into account expression (C.8) integral $I_T^{(f)}$ takes the form:

$$I_T^{(f)} = \frac{3\hbar^2}{8m\sigma_r^2} + \frac{\hbar^2}{2m\sigma_r^2}\int_0^\pi \frac{d\theta}{\sin\theta}. \tag{C.9}$$

Expression for integral $I_U^{(f)}$ coincides completely with $I_U^{(\mathrm{EM})}$. Indeed,

$$I_U^{(f)} = \frac{\hbar^2}{8m\sigma_r^4}\int_{\mathbb{R}^3} f(\vec{r})\left(r^2 - \frac{4\sigma_r^4}{r^2\sin^2\theta}\right)d^3r = I_U^{(\mathrm{EM})}. \tag{C.10}$$

Substituting (C.9) and (C.10) into expression (C.6), we obtain:

$$\langle\langle \mathcal{E} \rangle\rangle^{(f_w)} = \int_{\mathbb{R}^3}\int_{\mathbb{R}^3}\left(\frac{P^2}{2m} + U\right) f_w d^3r d^3P = \frac{3\hbar^2}{4m\sigma_r^2}. \tag{C.11}$$

Expressions (C.5) and (C.11) prove the validity of (1.34).